\documentclass[12pt]{article}
\usepackage{epsfig,amssymb,amsmath,psfrag,subfigure}
\usepackage{color,colordvi}
\usepackage{rotating}
\usepackage{ulem}

\usepackage{hyperref}

\catcode`\@=11
\def \tr {\mathop{\rm tr}\nolimits}

\def \e  {\mathop{\rm e}\nolimits}
\newcommand\lr[1]{{\left({#1}\right)}}
\newcommand \widebar [1] {\overline{#1}}

\newcommand\re[1]{(\ref{#1})}
\def \qqquad {\qquad\quad}
\def \qqqquad {\qquad\qquad}
 
\newcommand{\cM}{{\cal M}} 
\newcommand{\cF}{{\cal F}}
\newcommand{\cN}{{\cal N}}
\newcommand{\cA}{{\cal A}}

\newcommand{\cY}{{\cal Y}}

\newcommand{\nt}{\notag\\} 
\newcommand{\pa}{\partial}
\newcommand{\ep}{\epsilon}
\renewcommand{\a}{\alpha}
\renewcommand{\b}{\beta}
\newcommand{\g}{\gamma}
\newcommand{\da}{{\dot\alpha}}
\newcommand{\db}{{\dot\beta}}
\newcommand{\q}{\theta}
\newcommand{\bq}{\bar\theta}

\newcommand{\bx}{\bar\xi}

\newcommand{\ty}{\tilde{y}}
\newcommand{\cQ}{{\cal Q}}
\newcommand{\cI}{{\cal I}}
\newcommand{\cG}{{\cal G}}
\newcommand{\cT}{{\cal T}}
\newcommand{\px}[1]{\partial_{x_{#1}}}

\newcommand{\s}{\sigma}
\newcommand{\ts}{\tilde{\sigma}}

\newcommand{\p}[1]{(\ref{#1})}

\newcommand \vev [1] {\langle{#1}\rangle}
\newcommand{\ft}[2]{{\textstyle\frac{#1}{#2}}}

\newcommand{\OO}{O}

\newcommand{\rt}{\text{(0)}}
\newcommand{\nr}{\text{anom}}

\def\numberbysection{\@addtoreset{equation}{section}
                     \def\theequation{\thesection.\arabic{equation}}}

\numberbysection

\makeatletter
\def\timenow{\@tempcnta\time
  \@tempcntb\@tempcnta
  \divide\@tempcntb60
  \ifnum10>\@tempcntb0\fi\number\@tempcntb
  \multiply\@tempcntb60
  \advance\@tempcnta-\@tempcntb
  :\ifnum10>\@tempcnta0\fi\number\@tempcnta}
\makeatother

\begin{document}

\begin{titlepage}
 
\thispagestyle{empty}

\null\vskip-43pt \hfill
\begin{minipage}[t]{45mm}
CERN-PH-TH-2014-174\\
IPhT--T14--122\\
LAPTH--109/14
\end{minipage}

\vspace*{1cm}

\centerline{\large \bf  $\mathcal N=4$  superconformal Ward identities for correlation functions} 

\vspace*{1cm}

\centerline{\sc A.V.~Belitsky$^{a,b}$,  S.~Hohenegger$^c$, G.P.~Korchemsky$^b$, E.~Sokatchev$^{c,d,e}$ }

\vspace{5mm}

\centerline{\it $^a$Department of Physics, Arizona State University}
\centerline{\it Tempe, AZ 85287-1504, USA}

\vspace{3mm}

\centerline{\it $^b$Institut de Physique Th\'eorique\footnote{Unit\'e de Recherche Associ\'ee au CNRS URA 2306}, CEA Saclay}
\centerline{\it 91191 Gif-sur-Yvette Cedex, France}

\vspace{3mm}
\centerline{\it $^c$Physics Department, Theory Unit, CERN}
\centerline{\it CH -1211, Geneva 23, Switzerland}

\vspace{3mm}
\centerline{\it $^d$Institut Universitaire de France}
\centerline{\it  103, bd Saint-Michel
F-75005 Paris, France }

\vspace{3mm}
\centerline{\it $^e$LAPTH\,\footnote[2]{UMR 5108 du CNRS, associ\'ee \`a l'Universit\'e de Savoie},   Universit\'{e} de Savoie, CNRS}
\centerline{\it  B.P. 110,  F-74941 Annecy-le-Vieux, France}

\vspace{1cm}

\centerline{\bf Abstract}

\vspace{5mm}
\noindent
In this paper we study the four-point correlation function of the energy-momentum supermultiplet in theories with $\mathcal{N}=4$ superconformal symmetry in four dimensions. We present a compact form of all component correlators as an invariant of a particular abelian subalgebra of the $\mathcal{N}=4$ superconformal algebra. This invariant is unique up to a single function of the conformal cross-ratios which is fixed by comparison with the correlation function of the lowest half-BPS scalar operators. Our analysis is independent of the dynamics of a specific theory, in particular it is valid in $\mathcal{N}=4$ super Yang-Mills theory  for any value of the coupling constant.   We discuss in great detail a subclass of component correlators, which is a crucial ingredient for the recent study of charge-flow correlations in conformal field theories. We compute the latter explicitly and elucidate the origin of  the interesting relations among different types of flow correlations previously observed in arXiv:1309.1424.

\newpage
  
\end{titlepage}

\setcounter{footnote} 0

\newpage

\pagestyle{plain}
\setcounter{page} 1

\tableofcontents

\newpage 
 
\section{Introduction}

In this paper we study four-point correlation functions {involving conserved currents} in four-dimensional
theories with $\mathcal N=4$ superconformal symmetry. They include the $R-$symmetry current $J_\mu$,  
the supersymmetry currents $(\Psi^\a_{\mu},\bar \Psi^\da_{\mu})$ and the energy-momentum tensor $T_{\mu\nu}$.
These operators belong to the so-called $\mathcal N=4$ energy-momentum supermultiplet \cite{Bergshoeff:1980is} and  appear 
as various components in the expansion of the superfield $\mathcal T$ 
in powers of the 8 chiral ($\theta_\a^A$) and 8 antichiral ($\bar\theta^\da_A$) Grassmann variables, schematically, 
\begin{align}\label{T}
\mathcal T = O(x) +  (\theta \sigma^\mu \bar\theta) J_\mu(x)  + (\theta \sigma^\mu \bar\theta) \left[\q \Psi_{\mu}(x) + \bq \Psi_{\mu}(x)\right]   +(\theta \sigma^\mu \bar\theta) 
(\theta \sigma^\nu \bar\theta) T_{\mu\nu}(x) +  \dots\,.
\end{align}
Here the lowest component is a half-BPS scalar operator $O$ of {dimension} two, belonging
to the representation $\bf 20'$ of the $R-$symmetry group $SU(4)$.  The superfield \re{T} satisfies a half-BPS `shortness' condition, 
i.e., it is annihilated by half of the super-Poincar\'e generators. As a consequence, the expansion \re{T} is shorter than one would expect since 
$\mathcal T$ effectively depends on 4 chiral and 4 antichiral Grassmann variables only \cite{Howe:1981qj}. 

The central object of our study is the four-point correlation function of the energy-momentum supermultiplet  \re{T} in 
$\mathcal N=4$ superconformal theories. The most  widely studied example is  $\cN=4$ super-Yang-Mills theory (SYM) but in what follows we do not need to 
know any details about the dynamics of the theory. Our analysis is based solely on $\cN=4$ superconformal invariance and can be easily adapted to maximally supersymmetric theories in other space-time dimensions. 

$\cN=4$ superconformal 
symmetry is powerful enough to fix the form of the two- and three-point correlation functions of $\mathcal T$'s  \cite{D'Hoker:1998tz,Lee:1998bxa,Howe:1998zi}. In a perturbative theory, like $\cN=4$ SYM, the latter are protected from quantum corrections and only receive contributions at Born level~\cite{Penati:1999ba}. The four-point correlation function (we use the notation $(i)\equiv(x_i,\q_i,\bq_i)$)
\begin{align}\label{calG}
\mathcal G_4 = \vev{\mathcal T(1) \dots \mathcal T(4)} 
\end{align}
is the first and simplest example of an unprotected quantity.  It is this object that we study in the present paper. 

The super-correlation function \re{calG} combines together the correlation functions of various
components of the multiplet \re{T}. The latter appear as coefficients in the expansion of $\mathcal G_4$ in the
Grassmann variables. The lowest component of  $\mathcal G_4$ (with $\theta_i=\bar\theta_i=0$)
is the four-point correlation function of the half-BPS operators
\begin{align}\label{4O}
\mathcal G_4\big|_{\theta_i=\bar\theta_i=0} = \vev{ O(x_1) \dots  O(x_4)}\,.
\end{align}
$\cN=4$ superconformal symmetry fixes this correlation function up to a single function  $\Phi(u,v)$ of the two conformal cross-ratios $u$ and $v$~\cite{Eden:2000bk,Dolan:2004iy,Dolan:2004mu}. In the special case of $\mathcal N=4$ SYM, this function comprises the dependence on the coupling constant.  At weak coupling, its expansion
in terms of scalar conformal integrals has been worked out up to six loops~\cite{Eden:2011we,Eden:2012tu} and explicit expressions are available up to three loops 
\cite{GonzalezRey:1998tk,Eden:2000mv,Bianchi:2000hn,Drummond:2013nda}. At strong coupling, it has been
computed within the AdS/CFT correspondence in the supergravity approximation \cite{Arutyunov:1999fb,Arutyunov:2000py,Arutyunov:2000ku}. 

A unique feature of the super-correlation function \p{calG} is that the total number of Grassmann variables it depends upon 
(16 chiral and 16 antichiral variables)  matches the total number of  $\cN=4$ supercharges ($Q_\a^A$, $\bar Q^\da_A$, $S_\a^A$ and $\bar S^\da_A$). As we show below, this property alone ensures that
$\cN=4$ superconformal symmetry completely fixes all of its components, given the lowest one \p{4O}. The resulting expression 
for $\mathcal G_4$ has a remarkably simple form (see Eq.~\re{G4-main} below) and
is uniquely determined by the  function  $\Phi(u,v)$ that appears in the correlation function \re{4O}. 

The motivation for studying the correlation function \p{calG} is threefold. Firstly,
by performing the operator product expansion of $\mathcal G_4$
we can extract the spectrum of anomalous dimensions of various Wilson operators and evaluate the corresponding
three-point correlation functions. Both quantities are believed to be integrable in planar $\cN=4$ SYM \cite{Beisert:2010jr}.  

Secondly, a lot of attention has recently been devoted to the consistency conditions on the spectrum of conformal field theories \cite{Ferrara:1973yt,Polyakov:1974gs} that follow from the crossing symmetry of the four-point correlation functions
 (see \cite{ElShowk:2012ht} and references therein). This conformal bootstrap program was also extended to $\mathcal{N}=4$ superconformal  theories (see, e.g., \cite{Beem:2013qxa}), analyzing properties of the lowest component \p{4O}. The explicit 
expression for  the supercorrelator \p{calG} that we present in this paper might help in implementing supersymmetry more efficiently in 
 the analysis of the corresponding bootstrap equations.

Finally, various components of the supercorrelator \re{calG} can be used to compute interesting observables,
the so-called {\it charge-flow correlations}, measuring the flow
of various quantum numbers (e.g., $R-$charge, energy) in the final states created from a particular source.
They were first introduced in \cite{Ore:1979ry,Sveshnikov:1995vi,Korchemsky:1997sy,Korchemsky:1999kt,BelKorSte01}  in the context of  QCD, 
later  considered in \cite{Hofman:2008ar} in the context of $\cN=4$ SYM and recently studied in a systematical fashion in \cite{Belitsky:2013xxa,Belitsky:2013bja,Belitsky:2013ofa}.     
The charge-flow correlators  can be obtained from the four-point correlation function \re{calG} (after analytic continuation to Minkowski space) through 
a limiting procedure described in \cite{Sveshnikov:1995vi,Korchemsky:1997sy,Korchemsky:1999kt,BelKorSte01}. It involves sending two of the 
operators in \re{calG} to null infinity with a subsequent integration over their light-cone coordinates. 
As was noticed in \cite{Belitsky:2013xxa,Belitsky:2013bja,Belitsky:2013ofa}, for different choices of the quantum numbers,
the charge-flow correlations in $\cN=4$ SYM satisfy interesting relations (see Eq.~\p{univ} below), which hold for arbitrary coupling constant. 
In this paper, we elucidate the origin of these relations and show that they follow from the $\mathcal{N}=4$ superconformal symmetry of the correlation function \re{calG}.

The paper is organized as follows. In section 2, we formulate the conditions on the four-point correlation function \re{calG} imposed by $\cN=4$ 
superconformal symmetry and present a general solution to the corresponding Ward identities. In section 3, we examine the properties
of the various components of the super correlation function \re{calG} and work out a general expression for the correlation function involving
two scalar half-BPS operators and two conserved currents. We make use of this result in section 4 to establish relations between the charge-flow correlations 
previously observed in Ref.~\cite{Belitsky:2013bja}. Section 5 contains concluding remarks.  Some technical details are given in four appendices.

\section{Superconformal Ward identities}

In this section, we use $\cN=4$ superconformal symmetry to work out the general expression for the four-point correlation function \p{calG}
whose lowest component in the $\q$ and $\bq$ expansion is given by the correlation function of the half-BPS scalar operators \p{4O}.

\subsection{Lowest component}

The lowest component of the energy-momentum supermultiplet \re{T} is a scalar operator
 of conformal weight two. It belongs to the representation $\mathbf{20'}$ of the $R-$symmetry group 
 and carries two pairs of $SU(4)$ indices, $O_{\bf 20'} = O^{AB,CD}$.
For example, in $\cN=4$ SYM it is built from the six elementary scalar fields $\phi^{AB}(x)=-\phi^{BA}(x)$ (with $A,B=1,\dots,4$) and takes the form 
\begin{align}
O^{AB,CD}(x) =  \tr\left[\phi^{AB}(x) \phi^{CD}(x)\right] - \frac 1{12} \epsilon^{ABCD} \tr[\phi^{EF}(x)\phi_{EF}(x)]\,,
\end{align}
with $\phi_{EF} = \frac12 \epsilon_{EFKL}\phi^{KL}$. 

In order to keep track of the $SU(4)$ indices, it is convenient to project them with  auxiliary $SU(4)$ harmonic variables $u^{\pm a}_A$ introduced in Appendix~\ref{App:B} and parametrized  by analytic variables $y_{ab'}$ (see \re{3.17} in Appendix B) 
\begin{align}\label{O20}
& O (x,y) = O^{AB,CD}(x)\,Y_{AB}Y_{CD} \,,
\qquad
& Y_{AB}= u_A^{+c} \epsilon_{cd} u_B^{+d}  = \left[ \begin{array}{cc} \epsilon_{ab} & -y_{ab'}  \\  y_{ba'} & \epsilon_{a'b'} y^2 \end{array} \right]\,.
\end{align}
Here $y^2=\det y_{aa'}=\frac12 y_{aa'} \ty^{a'a}$ with $\ty^{a'a} =\epsilon^{ab}  y_{bb'}\epsilon^{b'a'}$
and we use composite indices $A=(a,a')$ (with $a,a'=1,2$) and similarly for $B=(b,b')$. Notice that the operator \p{O20} is
quadratic in the (isotopic) $Y-$variables.

The meaning of the variables $y_{ab'}$ is very similar to that of the space-time coordinates $x_{\a\db}$. They parametrize holomorphic compact four-dimensional cosets of the $R-$symmetry (for $y$) and of the conformal (for $x$) groups, or more precisely, of their complexification  $GL(4,\mathbb{C})$ (see \p{cos} in Appendix B). 
In this way, the $R-$symmetry and conformal groups are realized as $SL(4,\mathbb{C})$ transformations of the coordinates $y$ or $x$, respectively. A particularly useful subgroup is generated by  translations and inversion of the coordinates $x$ or $y$. Combining translations with inversions, one can obtain any element of the group.  The half-BPS operator \p{O20} transforms covariantly under inversions of $x$ and $y$  with weight $(-2)$ and $(+2)$, respectively
(see Eqs.~\p{in1} and \p{in2} below).

The correlation functions of the half-BPS operators \re{O20} have been systematically studied in the literature \cite{GonzalezRey:1998tk,Eden:1998hh,Eden:1999kh,Eden:2000bk,Eden:2000mv,Bianchi:2000hn}.  Below we briefly review  the properties of the Euclidean four-point correlation function 
 \begin{align}
G_4\equiv \mathcal G_4\big|_{\theta_i=\bar\theta_i=0} =\langle O(x_1,y_1) O(x_2,y_2) O(x_3,y_3) O(x_4,y_4) \rangle\,.\label{Scalar4ptCorr}
\end{align}
As a function of the space-time intervals $x_{ij}^2=(x_i-x_j)^2$, it can be decomposed into two parts, 
\begin{align}\label{rnr}
G_4=G_4^{\rt}+G_4^{\nr}\, .
\end{align}
The difference between them can be seen when performing an OPE decomposition of the four-point function and
examining the resulting short distance asymptotics of $G_4$ for $x_{ij}^2\to 0$.
The first part, $G_4^{\rt}$, receives contributions from operators with canonical conformal weights. As a consequence, $G_4^{\rt}$ is a rational function 
of $x_{ij}^2$. The second part, $G_4^{\nr}$, 
receives contributions from unprotected operators and its asymptotic behavior for $x_{ij}^2\to 0$ is controlled by the anomalous conformal weights of these operators. 

For example, in perturbative $\cN=4$ SYM the first part describes the Born approximation 
to $G_4$ and is given by a product of free scalar propagators dressed with 
$y-$dependent factors,  
\begin{align}\notag\label{G-Born}
G_4^{\rt}= {} &  \frac{y^2_{12}}{ x^2_{12}} \frac{y^2_{23}}{ x^2_{23}} \frac{y^2_{34}}{ x^2_{34}} \frac{y^2_{14}}{ x^2_{14}}  
+\frac{y^2_{13}}{ x^2_{13}} \frac{y^2_{23}}{ x^2_{23}} \frac{y^2_{24}}{ x^2_{24}} \frac{y^2_{14}}{ x^2_{14}}
+ \frac{y^2_{12}}{ x^2_{12}} \frac{y^2_{24}}{ x^2_{24}} \frac{y^2_{34}}{ x^2_{34}} \frac{y^2_{13}}{ x^2_{13}}  
\\{}
+ {}&{1\over 4} (N_c^2-1)  \left[\lr{\frac{y^2_{12}}{ x^2_{12}} \frac{y^2_{34}}{ x^2_{34}}}^2  + \lr{\frac{y^2_{13}}{ x^2_{13}} \frac{y^2_{24}}{ x^2_{24}}}^2  + \lr{\frac{y^2_{14}}{ x^2_{14}} \frac{y^2_{23}}{ x^2_{23}}}^2\right]\,,
\end{align}
where 
$x_{ij}^2=(x_i-x_j)^2$ and 
$y_{ij}^2=(y_i-y_j)^2$, and $N_c$ refers to the gauge group $SU(N_c)$.
The second part encodes the perturbative  corrections to $G_4$,
\begin{align}\nonumber\label{G4-loop}
G_4^{\nr}
=  & \bigg[  \frac{y_{12}^2y_{23}^2y_{34}^2y_{41}^2}{x_{12}^2x_{23}^2x_{34}^2x_{41}^2}(1-u-v)+\frac{y_{12}^2y_{13}^2y_{24}^2y_{34}^2}{x_{12}^2x_{13}^2x_{24}^2x_{34}^2}(v-u-1)+   \frac{y_{13}^2y_{14}^2y_{23}^2y_{24}^2}{x_{13}^2x_{14}^2x_{23}^2x_{24}^2}(u-v-1)
\\[2mm]
&+\lr{\frac{y^2_{12}}{ x^2_{12}} \frac{y^2_{34}}{ x^2_{34}}}^2u+ \lr{\frac{y^2_{13}}{ x^2_{13}} \frac{y^2_{24}}{ x^2_{24}}}^2  + \lr{\frac{y^2_{14}}{ x^2_{14}} \frac{y^2_{23}}{ x^2_{23}}}^2v \bigg]\Phi(u,v)\,,
\end{align}
where the scalar function $\Phi(u,v)$ depends on the parameters of the theory (the coupling constant,  the gauge group Casimirs, etc.) and  on the two conformal cross-ratios
\begin{align}\label{uv}
u={ x_{12}^2 x_{34}^2 \over x_{13}^2 x_{24}^2} \,,\qqqquad v={ x_{23}^2 x_{41}^2 \over x_{13}^2 x_{24}^2}\,.
\end{align}
The Bose symmetry of the correlation function \re{G4-loop} leads to the crossing symmetry relations 
\begin{align}\label{cross}
 \Phi(u,v) = \frac1{v} \Phi(u/v,1/v)= \frac1{u} \Phi(1/u,v/u)\,.
\end{align}

The specific polynomial in the square brackets in \p{G4-loop} is universal and does not depend on dynamical details of the theory. Its presence is a corollary of $\cN=4$ conformal supersymmetry and the requirement of polynomial dependence on the auxiliary coordinates $y$ (see, e.g.,  \cite{Dolan:2004mu,Heslop:2002hp} and Section~\ref{s2.4}). It can be rewritten in a more compact form,
\begin{align}\label{match'}
G_4^{\nr} = \lr{\frac{y^2_{13}}{x^2_{13}}\frac{y^2_{24}}{x^2_{24}}}^2
 \ (\zeta-w)(\zeta-\bar w)(\bar \zeta-w)(\bar \zeta-\bar w)\, \frac{\Phi(u,v)}{uv}\,,
\end{align}
where the new variables $\zeta$ and $\bar \zeta$ are defined by means of the relations  $\zeta\bar \zeta=u$ and $(1-\zeta)(1-\bar \zeta) =v$, and  
the variables $w$ and $\bar w$ are defined through  the analogous $SU(4)$  cross-ratios,
\begin{align}\label{210} 
 U={y_{12}^2 y_{34}^2\over y_{13}^2y_{24}^2} = w\bar w\,,\qqqquad  V={y_{23}^2 y_{14}^2\over y_{13}^2y_{24}^2} = (1-w)(1-\bar w)\,.
\end{align}	

As already mentioned before, the function $\Phi(u,v)$ is known at weak coupling up to six loops~\cite{Eden:2011we,Eden:2012tu} in terms of conformally 
covariant scalar Feynman integrals~\cite{Usyukina:1992wz}
and has also been computed at strong coupling using the AdS/CFT correspondence~\cite{Arutyunov:1999fb,Arutyunov:2000py,Arutyunov:2000ku}. 
The short distance asymptotics of this function for $x_{12}^2\to 0$ goes in  powers of $\ln u$ and $(1-v)$, thus giving rise to anomalous contributions 
to the conformal weights of operators (see, e.g., \cite{Dolan:2004iy} for details). 
 
\subsection{Higher components} \label{desc}

We recall that the BPS operator \re{O20} is annihilated by half of the Poincar\'e supercharges.
Using harmonic variables (see Appendix \ref{App:B}), this half
corresponds to the projections $Q_{-a'}^\alpha= \bar u^A_{-a'} Q_A^\alpha$ and $ \bar Q^{\dot\alpha , {+a}} = u^{+a}_A \bar Q^A_{\dot\alpha}$,  
\begin{align}
Q_{-a'}^\alpha O(x,y) = \bar Q^{\dot\alpha, {+a}}  O(x,y) =0\,.
\end{align}
We can use the remaining supercharges, $Q_{+a}^\alpha= \bar u^A_{+a} Q_A^\alpha$ and $ \bar Q_{\dot\alpha}^{-a'} = u^{-a'}_A \bar Q^A_{\dot\alpha}$ to construct the energy-momentum supermultiplet 
\begin{align}\label{T-gen0}
\mathcal T(x,\theta,\bar\theta,y) =  \exp\lr{\theta^a_{\alpha} Q_{+a}^\alpha + \bar\theta^{\dot\alpha }_{a'}  \bar Q_{\dot\alpha}^{-a'}} O(x,y)\,.
\end{align}
It depends on half of the $\cN=4$ Grassmann variables, four chiral $\theta^a_{\alpha}$ and four antichiral $\bar\theta_{\dot\alpha a'}$. 
Expanding \re{T-gen0} in powers of the latter and applying $\cN=4$ supersymmetry  transformations to the scalar fields in $O(x,y)$, we can work out explicit expressions for the
various components of the energy-momentum supermultiplet.\footnote{The chiral part of the supermultiplet corresponding to 
$\mathcal T(x,\theta,\bar\theta=0,y)$ can be found in Ref.~\cite{Eden:2011yp}.} In this way, we find  
\begin{align}\label{T-gen}
\mathcal T(x,\theta,\bar\theta,y) = O(x,y) + \theta^{\alpha a}  \bar \theta^{\dot\alpha a'}  \widehat J_{\alpha\dot\alpha,aa'}(x,y)+\dots\,,
\end{align}
where the component  $\widehat J_{\alpha\dot\alpha,aa'}(x,y)$ can be written as a sum of two terms,
\begin{align}\label{J}
\widehat J_{\alpha\dot\alpha,aa'}(x,y)  =   J_{\alpha\dot\alpha,aa'}(x,y)-\frac12 \frac{\partial}{\partial x^{\dot\alpha\alpha}}
\frac{\partial}{\partial y^{a'a}} O(x,y)\,.
\end{align}
Here the first term on the right-hand side is the conserved $R-$symmetry current $(J_{\a\da})_B^A$  projected with two
harmonic matrices,  $J_{\alpha\dot\alpha}{{}^a}_{a'}(x,y) = (J_{\a\da})_B^A u_{A}^{+a} \bar u^B_{- a'}$  (with $(J_{\a\da})^A_A=0$), transforming in the $\mathbf{15}$ of $SU(4)$. The second, derivative term in \p{J} is a descendant of the half-BPS scalar operator $O(x,y)$ with respect to both the conformal and $R-$symmetry groups. 

We deduce from  \re{T-gen} and \re{J} that the $R-$symmetry current  can be extracted from the superfield $\mathcal T(x,\theta,\bar\theta,y)$ by
applying the following differential operator
\begin{align}\label{diffJ}
J_{\alpha\dot\alpha,aa'}(x,y) =\left[ (\partial_{\bar\theta})_{\dot\alpha a'}(\partial_{\theta})_{\alpha a} 
+
\frac12 (\partial_x)_{\alpha\dot\alpha} 
(\partial_y)_{aa'}\right] \mathcal T(x,\theta,\bar\theta,y)\bigg|_{\theta=\bar\theta=0}\,,
\end{align}
where $(\partial_{\bar\theta})_{\dot\alpha a'}=\partial/\partial \bq^{\dot\alpha a'}$ and $(\partial_{\theta})_{\alpha a}=\partial/\partial \q^{\alpha a}$.
Here the relative coefficient $1/2$ is needed for the current conservation condition, $(\partial_x)^{\dot\alpha \alpha} J_{\alpha\dot\alpha,aa'}(x)=0$. 

The same pattern also holds for the higher components of the expansion shown in \re{T-gen}. Namely, they are given by a linear combination
of conserved currents and total derivatives acting on lower components of the superfield. Like in \re{diffJ}, the energy-momentum
tensor can be extracted from $\mathcal T$ with the help of a differential operator, namely, 
\begin{align}\notag\label{diffT}
& T_{\alpha\dot{\alpha},\beta\dot{\beta}}(x) = 
\left[  - (\partial_{\q})_{\a}^{a}   (\partial_{\q})_{ \b a} (\partial_{\bq})_{\da a'} (\partial_{\bq})_{\db}^{a'}  
-
(\partial_{\q})_{(\a}^{a} (\partial_{x})_{\b)(\db}(\partial_{y})_{aa'} (\partial_{\bq})_{\da)}^{a'} \right.
\\ 
 &\hspace*{37mm} 
 \left. 
+
 \frac1{6}  (\partial_{x}) _{(\a\da} (\partial_{x}) _{\b)\db}
 (\partial_{y})_{aa'} (\partial_{y})^{a'a}\right]\mathcal T(x,\theta,\bar\theta,y)\big|_{\theta=\bar\theta=0}\,,
\end{align}
where $(\a\b)$ denotes weighted symmetrization, i.e, $(\alpha\beta) = \ft12 (\alpha\beta + \beta\alpha)$.
As before, the relative coefficients in this expression ensure the conservation of the energy-momentum tensor 
$(\partial_x)^{\dot\alpha \alpha}  T_{\alpha\dot{\alpha},\beta\dot{\beta}} = 0$, and, in addition, the independence of $ T_{\alpha\dot{\alpha},\beta\dot{\beta}}$ of the harmonic variables $y$. The latter follows from the fact that the energy-momentum tensor is 
an $SU(4)$ singlet. In the same fashion, the conserved supersymmetry current $\Psi_{\alpha\beta\dot{\alpha},a'}(x,y)  = \Psi_{\alpha\beta\dot{\alpha}\ A}(x)\, \bar u^A_{-a'}$, 
which transforms in the fundamental representation of $SU(4)$, is extracted with the help of the operator
\begin{align}\label{diffS}
\Psi_{\alpha\beta\dot{\alpha},a'}(x,y)  =  
\left[ (\partial_{\theta})_{\a a}(\partial_{\theta})_{\b}^a(\partial_{\bar{\theta}})_{\da a'}
+ \frac{2}{3}
(\partial_x)_{(\a\da}(\partial_y)_{a' a} (\partial_{\theta})_{\b) }^a \right]\mathcal T(x,\theta,\bar\theta,y)\big|_{\theta=\bar\theta=0}\,.
\end{align}

Relations \re{diffJ}, \re{diffT} and \p{diffS} imply that the correlation functions involving conserved currents can be obtained
from $\mathcal G_4$ defined in \p{calG}  using  corresponding differential operators, for instance,
 \begin{align}\label{JOOO}
\vev{ J_{\alpha\dot\alpha,aa'}(1)\, O(2)\, O(3) \,O(4)} =  \left[ (\partial_{\bar\theta_1})_{\dot\alpha a'}(\partial_{\theta_1})_{\alpha a} + \frac12 (\partial_{x_1})_{\alpha\dot\alpha} 
(\partial_{y_1})_{a'a}\right] \mathcal G_4\bigg|_{\theta=\bar\theta=0} \,.
\end{align}
We would like to emphasize that these differential operators have a purely kinematical origin. They do not depend on the details of the dynamics, e.g., on the coupling constant. The latter is encoded in the correlation function  $\mathcal G_4$, more precisely, in the function $\Phi(u,v)$ in \p{G4-loop}. 
 
\subsection{$\cN=4$ superconformal transformations}  

By construction, the correlation function of  the superfields $\mathcal T(x,\theta,\bar\theta,y)$ should be covariant under
$\cN=4$ superconformal  transformations. 

As we explain below, to construct the four-point correlation function $\mathcal G_4$
we only need to examine the action of half of the supersymmetry generators, namely  the chiral Poincar\'e supercharges $Q$ and the antichiral
special superconformal transformations $\bar S$ \footnote{The latter can be realized as a composition of inversion $I$ and supersymmetry
transformations,  $\bar S= I \cdot Q \cdot I$. } 
\begin{align}\label{varT}\notag
\mathcal T{\phantom{'}}'(x,\theta,\bar\theta,y) &{} =\e^{(\epsilon\cdot Q)+ (\bx \cdot \bar S)} \mathcal T(x,\theta,\bar\theta,y)
\\[2mm]
&{}
=(1+\delta w)  \mathcal T(x+\delta x,\theta+\delta \theta,\bar\theta+\delta \bar\theta,y+\delta y)\,,
\end{align}
where we used a shorthand notation for $(\epsilon\cdot Q)=\epsilon_\a^{ A} Q^\a_{A}$ and $(\bx \cdot \bar S)= \bx^{\da A}  \bar S_{\da A}$. The  infinitesimal transformations of the supercoordinates are
\begin{align}\label{CSUSY}\notag
& \delta x_{\a\da}  = (\ep^{a'}_\a + x_{\a\db} \bx^{\db a'}) \bq_{a'\da}\,,  && \delta y_{a'}{}^a  =  \bq_{a' \db} (\bx^{\db a}+\bx^{\db b'}  y_{b'}{}^a)\,,
\\[2mm]
& \delta \q^{a}_\a  = \ep^{a}_\a +\ep^{a'}_\a y_{a'}{}^a + x_{\a\db}(\bx^{\db a}+\bx^{\db a'}  y_{a'}{}^a)\,, &&
\delta \bq_{a'\da}  =   \bx^{\db b'} \bq_{b'\da} \bq_{a'\db} \,,
\end{align}
where $\epsilon_\alpha^A=(\epsilon_\alpha^{a},\epsilon_\alpha^{a'})$ and $ \bx^{\da A}=( \bx^{\da a}, \bx^{\da a'})$ are the parameters 
of the $Q-$ and $\bar S-$transformations respectively. The weight $\delta w$    in \re{varT} reflects  the  conformal weight of the half-BPS operator \re{O20} 
\begin{align}\label{w0}
\delta w =  2 \,  \bx^{\db b'} \bq_{b'\db} \,.
\end{align} 
The  generators of the transformations \p{CSUSY} are given by linear differential operators
\begin{align}\notag\label{generators}
& Q^\alpha_a = {\partial \over \partial \q_\alpha^a}\,,\hspace{1.5cm} 
\\\notag
& Q^\alpha_{a'} = \bq_{a'\dot\alpha} {\partial\over\partial x_{\alpha\dot\alpha}} + y_{a'}{}^a {\partial\over\partial \theta_{\alpha}^a}\,,\hspace{1.5cm}
\\\notag
&
 \bar S_{a\dot\beta} = -\bq_{a'\dot\beta} {\partial \over \partial y_{a'}{}^a}  + x_{\alpha\dot\beta}  {\partial \over \partial \q_\alpha^a}\,,
\\
& \bar S_{b'\dot\beta} = x_{\alpha\dot\beta} \bq_{b'\dot\alpha} {\partial\over\partial x_{\alpha\dot\alpha}} +  x_{\alpha\dot\beta} y_{b'}{}^a 
{\partial\over\partial \theta_{\alpha}^a}- \bq_{a'\dot\beta} y_{b'}{}^a  {\partial\over\partial y_{a'}{}^a} + \bq_{b'\dot\alpha} \bq_{a'\dot\beta}
{\partial\over\partial \bq_{a' \dot\alpha}}\,.
\end{align} 
It is easy to check that they form an abelian subalgebra, i.e.  
$\{Q,Q\}=\{Q,\bar S\}= \{\bar S, \bar S\}=0$.

The  correlation functions of the supermultiplets  $\mathcal T$ should be covariant under
the transformations \re{varT}. This leads to the following {superconformal Ward identity} for the four-point correlation function $\mathcal G_4$ 
\begin{align}\label{WI}
\mathcal G_4\left(x_i,\q_i,\bq_i,y_i\right) = \big(1+\sum_{k} \delta w_k\big) \mathcal G_4\left(x_i+\delta x_i,\q_i+\delta\q_i,\bq_i+\delta\bq_i,y_i+\delta y_i\right)\,.
\end{align}
Substituting \p{CSUSY} and \p{w0} into this relation and comparing the coefficients in front of $\epsilon$ and $\bx$, we 
obtain a system of linear differential equations for $\mathcal G_4$. Combining it with the expressions for the lowest
component of the supercorrelator, Eqs.~\p{Scalar4ptCorr} and \p{rnr}, we can determine  $\mathcal G_4$.
Similarly to \p{rnr},  the general solution to \p{WI} can be decomposed into rational and anomalous parts,
\begin{align}
\mathcal G_4=\mathcal G_4^{\rt}+ \mathcal G_4^{\nr}\,,
\end{align}
where the lowest components of $\mathcal G_4^{\rt}$ and $\mathcal G_4^{\nr}$ are given by \p{G-Born} and \p{G4-loop}, 
respectively.

\subsection{Uniqueness of the superconformal extension}\label{s24}

The four-point correlation functions of half-BPS short supermultiplets like the energy-momentum multiplet \p{T} have a very important property. Their half-BPS nature guarantees the uniqueness of the superconformal extension \p{calG} of the lowest component \p{4O} \cite{Eden:2000gg}. The underlying reason for this is based on a simple counting of the Grassmann degrees of freedom. Each operator $\cT$ in \p{calG} depends on 4 chiral odd variables $\q^a_\a$ and on 4 antichiral ones $\bq_{a'\a}$. Altogether the four-point function depends on 16 chiral and 16 antichiral variables. Further, the 16 generators  $Q$ and $\bar S$ in \p{generators} act, essentially, as shifts of the chiral odd variables $\q^a_\a$, and similarly for the conjugates $\bar Q$ and $S$ with respect to the antichiral $\bq_{a'\a}$. This implies that by making a finite $\cN=4$ superconformal transformation~\footnote{However, its practical realization is non-trivial, due to the non-Abelian nature of the superconformal algebra.}
we can fix a frame in which all $\q_i=\bq_i=0$ for $i=1,2,3,4$. In such a frame the supercorrelator \p{calG} is reduced to its lowest component \p{4O}. Inversely, starting from the latter and making the same finite $\cN=4$ superconformal transformations
we can restore the dependence on $\q_i$ and $\bq_i$ in a {\it unique way}.\footnote{The same counting argument applies to two- and three-point functions of half-BPS short supermultiplets.} 

The uniqueness of this supercorrelator can also be understood in the following way. Suppose that there exist two different supercorrelators $\cG_4$ and $\cG_4'$  which share the same lowest component \p{4O}. Then their difference $\cG_4-\cG_4'$ would be a {nilpotent} superconformal covariant proportional to the odd variables. Consequently, in the fixed frame  $\q_i=\bq_i=0$ this difference vanishes and so it must vanish in any frame by virtue of superconformal covariance. 

This very special property of the four-point correlation function of the energy-momentum supermultiplet \p{T} allows us to develop a strategy for reconstructing it from its lowest component \p{4O}. The strategy is different for the rational and anomalous parts in \p{rnr} and is explained in detail below.

\subsection{Rational part}

The reconstruction of the rational part $\mathcal G_4^{\rt}$ of the supercorrelator relies on a simple observation 
that its lowest component \p{G-Born} is given by a product of free scalar propagators $y_{ij}^2/x_{ij}^2$. This
suggests that  $\mathcal G_4^{\rt}$ can be obtained from $G^{\rt}_4$ in \p{G-Born} by replacing the free scalar
propagators by their $Q-$supersymmetric version~\cite{Heslop:2002hp} \footnote{The two-point function of a half-BPS short supermultiplet is overdetermined, just like the two-point function of a conformal scalar filed is determined by translations and dilatation alone. So, the form of $\hat x$ is fixed by the requirement that the two-point function (propagator) be invariant under the $Q-$supersymmetry generators from \p{generators} and their conjugates $\bar Q$.}
\begin{align}\label{susy-pro}
{y_{ij}^2 \over x_{ij}^2} \ \ \to \ \  {y_{ij}^2 \over \hat x_{ij}^2}\,,\qqqquad \hat x_{ij}^{\dot\alpha\alpha} = x_{ij}^{\dot\alpha\alpha}  - \q^{a\alpha}_{ij} (y^{-1}_{ij})_{aa'} \bq^{a'\dot\alpha}_{ij}\,,
\end{align}
with $\q_{ij}=\q_i-\q_j$ and $y_{ij}=y_i-y_j$.   The supersymmetrized scalar propagator defined in this way is automatically covariant under the rest of the $\cN=$ superconformal 
algebra, in particular, under $\bar S_{a'\da}$,
\begin{align}\label{w}
\delta_{\bar S} \frac{y^2_{ij}}{\hat x^2_{ij}} =   \bx^{\da a'}(\bq_i+\bq_j)_{\da a'} \frac{y^2_{ij}}{\hat x^2_{12}}\,.  
\end{align}
Then, supersymmetrizing the propagators in   \re{G-Born} according to \p{susy-pro} we find  
\begin{align}\notag\label{calG-Born}
\mathcal G_4^{\rt}= {} &   \left(\frac{y^2_{12}}{ \hat x^2_{12}} \frac{y^2_{23}}{\hat  x^2_{23}} \frac{y^2_{34}}{\hat  x^2_{34}} \frac{y^2_{14}}{ \hat x^2_{14}}  
+\frac{y^2_{13}}{\hat  x^2_{13}} \frac{y^2_{23}}{\hat  x^2_{23}} \frac{y^2_{24}}{\hat  x^2_{24}} \frac{y^2_{14}}{\hat  x^2_{14}}
+ \frac{y^2_{12}}{\hat  x^2_{12}} \frac{y^2_{24}}{ \hat x^2_{24}} \frac{y^2_{34}}{\hat  x^2_{34}} \frac{y^2_{13}}{ \hat x^2_{13}}  \right) 
\\{}
+ {}&{}  {1 \over 4} (N_c^2-1)\left[\lr{\frac{y^2_{12}}{ \hat x^2_{12}} \frac{y^2_{34}}{\hat  x^2_{34}}}^2  + \lr{\frac{y^2_{13}}{\hat  x^2_{13}} \frac{y^2_{24}}{\hat  x^2_{24}}}^2  + \lr{\frac{y^2_{14}}{\hat  x^2_{14}} \frac{y^2_{23}}{\hat  x^2_{23}}}^2\right],
\end{align}
where the dependence on the Grassmann variables resides only in  $\hat x_{ij}^2$. We can use the transformation rule \p{w} to verify that  \re{calG-Born} indeed satisfies the superconformal Ward identity \re{WI}. As explained in Section~\ref{s24}, the superconformal extension \p{calG-Born} is unique.

The expansion of \re{calG-Born} in powers of the Grassmann variables produces the Born level approximation to the correlation functions of the various
components of the energy-momentum supermultiplet in $\cN=4$ SYM. For instance, substituting \re{calG-Born} into \re{JOOO}, we find after some algebra \footnote{Notice that the second term in \re{calG-Born} does not contribute because it factorizes into $\vev{J_{\alpha\dot\alpha,aa'}(1) \OO(2)}\vev{\OO(3) \OO(4)}$ but the first factor vanishes in virtue of conformal and $R-$symmetry.}
\begin{align}\label{onecu}\notag
& \vev{ J_{\alpha\dot\alpha,aa'}(1)\, O(2)\, O(3) \,O(4)}^{\rt}  = \frac1{2 {x^2_{12} x^2_{23} x^2_{34} x^2_{14}}}  
\\[2mm] 
& \qquad \times \big[ {y^2_{23} y^2_{34}} (Y_{124})_{aa'}(X_{124})_{\alpha\dot\alpha}  + {y^2_{23} y^2_{24}} (Y_{134})_{aa'} u (X_{134})_{\alpha\dot\alpha} + {y^2_{24} y^2_{34}} (Y_{123})_{aa'} v (X_{123})_{\alpha\dot\alpha}\big]  \,,
\end{align}
where the conformal cross-ratios $u$ and $v$ were defined in \re{uv}. The three-point $x-$ and $y-$dependent
structures
\begin{align}\label{XY}
&X_{ijk} = x_{ij}^{-1} x_{jk} x_{ki}^{-1} \,,\qqqquad Y_{ijk} = y_{ij} y_{jk} y_{ki} 
\end{align}
are covariant under conformal  and   $SU(4)$ transformations, respectively.\footnote{We recall that the $y$'s are coordinates on a coset  of $SU(4) $ 
(or rather its  complexification $GL(4,\mathbb{C})$), just like the $x$'s are coordinates on a coset  of  $SU(2,2)$.} The simplest way to check this is to 
perform an inversion in the $x-$space, $x_{\a\da}\to x^{\da\a}/x^2$, and an analogous inversion in the $y-$space, $y_{ab'} \to y^{b'a}/y^2$, 
\begin{align}\label{YI}
&X_{ijk} \ \stackrel{I_x}{\rightarrow} \ x_i X_{ijk} x_i\,,\qqqquad Y_{ijk} \stackrel{I_y}{\longrightarrow} \ y_i Y_{ijk} y_i/(y^{4}_i  y^{2}_j y^{2}_k)\,. 
\end{align}
This shows that $X_{ijk}$ has conformal weight $(-1)$ at point $x_i$ and weight zero at points $x_j$ and $x_k$, while $Y_{ijk}$ has $SU(4)$ weights 
at the three points. {The conserved current $J_{\a\da,aa'}$ and the half-BPS scalar operator $O$ are conformal primaries transforming under inversion 
in the $x-$space according to their weights:
\begin{align}\label{in1}
&J_{\a\da,aa'} \ \stackrel{I_x}{\rightarrow} \ {(x^2)^2} x^{\da\b} J_{\b\db,aa'} x^{\db\a}\,, \qqqquad O \ \stackrel{I_x}{\rightarrow} \ {(x^2)^2}  O\,.
\end{align}
Similarly, $J_{\a\da,aa'}$ and $O$ belong to the representations $\mathbf{15}=[1,0,1]$ and $\mathbf{20'}=[0,2,0]$ of  $SU(4)$, respectively, and 
transform under  inversion in the $y-$space as \footnote{The reader will remark the similarity between \p{in1} and \p{in2}. The only difference 
between theses conformal and $R-$symmetry representations is in the sign of the weight, which leads to infinite and finite dimensional unitary irreps, respectively.}
\begin{align}\label{in2}
&J_{\a\da,aa'} \ \stackrel{I_y}{\rightarrow} \ (y^2)^{-2} y^{a'b} J_{\a\da,bb'} y^{b'a}\,,\qqqquad O \ \stackrel{I_y}{\rightarrow} \ (y^2)^{-2} O\,.   
\end{align}
It is easy to check that the expression \re{onecu} has the correct conformal and $SU(4)$ transformation properties. In addition, it 
is straightforward to verify that the correlation function \re{onecu} vanishes upon the action of the derivative $(\partial_{x_1})^{\dot\alpha\alpha}$ 
in accord with the current conservation $(\partial_{x_1})^{\dot\alpha\alpha} J_{\alpha\dot\alpha,aa'}(x_1)=0$. }\footnote{We neglect here contact terms.}

\subsection{Anomalous part}\label{s26}

The supersymmetrization of the anomalous part of the correlator \p{rnr} is more elaborate since \p{G4-loop} involves a nontrivial
function $\Phi(u,v)$ and, therefore, is not reduced to a product of free scalar propagators.

Let us rewrite the anomalous contribution $\mathcal G^{\nr}_4$ by pulling out  a propagator factor,   
\begin{align}\label{calG4-loop}
\mathcal G_4^{\nr}= \lr{\frac{y^2_{13}}{\hat x^2_{13}}\frac{y^2_{24}}{\hat x^2_{24}}}^2  \mathcal I_4\left(x_i,\q_i,\bq_i,y_i\right) \,.
\end{align}
{This factor carries the necessary conformal and $SU(4)$ weights of the correlation function. } Substituting \re{calG4-loop} into the Ward 
identity \re{WI} and taking into account \re{w}, we find that the function $\mathcal I_4$ should be invariant under the superconformal
transformations \re{CSUSY},
\begin{align}
\mathcal I_4\left(x_i,\q_i,\bq_i,y_i\right) =  \mathcal I_4\left(x_i+\delta x_i,\q_i+\delta\q_i,\bq_i+\delta\bq_i,y_i+\delta y_i\right)\,,
\end{align} 
or equivalently,
\begin{align}\label{I-inv}
 Q_{A}^\alpha\, \mathcal I_4\left(x_i,\q_i,\bq_i,y_i\right) = \bar S_{\da A}\, \mathcal I_4\left(x_i,\q_i,\bq_i,y_i\right) =0\,.
\end{align}  
Here the generators $Q_{A}^\alpha$ and $\bar S_{\da A}$ (with $A=(a,a')$) are given by the differential operators \re{generators} acting 
on the coordinates of the four points. Likewise, the invariance of $\mathcal I_4$ under the second half of the
odd generators leads to 
\begin{align}\label{I-inv1}
  \bar Q_{\da}^A\, \mathcal I_4\left(x_i,\q_i,\bq_i,y_i\right) =S^{\a A}\, \mathcal I_4\left(x_i,\q_i,\bq_i,y_i\right) =0\,.
\end{align}

We now apply the argument of Section~\ref{s24} to the invariant $\mathcal I_4\left(x_i,\q_i,\bq_i,y_i\right)$ in order to reconstruct it starting from the lowest component \p{G4-loop}. We will do this in three steps: (i) solve only the first half \p{I-inv} of the Ward identities; (ii) match the solution with the lowest component \p{G4-loop}; (iii) show that, under certain conditions, the solution of \p{I-inv} also automatically solves the other half \p{I-inv1}. In this way we can be sure to have constructed the full and unique supersymmetric extension of \p{G4-loop}.

We recall that the generators  $Q$ and $\bar S$ are nilpotent, $(Q_{A}^\alpha)^2=(\bar S_{\da A})^2=0$ and that they anti commute, $\{Q, \bar S\}=0$. This suggests to write the  general solution to \re{I-inv} in the form
\begin{align}\label{calI}
 \mathcal I_4\left(x_i,\q_i,\bq_i,y_i\right) = Q^4 Q'{}^4 \bar S^4 \bar S'{}^4 \mathcal A_4(x_i,\q_i,\bq_i,y_i)
\end{align}
with  an arbitrary function $\mathcal A_4$.
Here we employed the conventions from Appendix \ref{AppendixNotations} such that
\begin{align}\label{Qdef}
Q^4= \frac{1}{12} Q_{a}^\alpha Q_\a^b Q_b^\b Q_\b^a
\, , \qquad\qquad
Q'{}^4 = \frac{1}{12} Q^{a' \alpha} Q_{b'\alpha} Q^{b'\beta} Q_{a' \beta}
\, ,
\end{align}
with a natural contraction of (un)dotted and (un)primed indices and similarly for the rest, i.e., $\bar S^4$ and $\bar S'{}^4$. The  normalization  
in the above equations was chosen so that  $Q^4 = Q_1^1 Q_2^1 Q_1^2 Q_2^2 $, etc.

To determine the function $\mathcal A_4$ we examine \p{calI} in the chiral sector $\bar\theta_i=0$. We observe from  \p{generators}  that for $\bar\theta_i=0$  the generators $Q$ and $\bar S$ are reduced to partial derivatives $\pa/\pa\q_i$. 
The corresponding part of  the differential operator in \p{calI} removes all $\q^{a\a}_i$ (with $i=1,\dots,4$) from $\mathcal A_4$. Therefore, for \p{calI} to
be different from zero if all $\bar\theta_i=0$, $\mathcal A_4$ should be proportional to the product of all $\theta$'s
\begin{align}\label{calA}
 \mathcal A_4(x_i,\q_i,\bq_i,y_i) = \q_1^4 \, \q_2^4 \, \q_3^4 \, \q_4^4 A_4(x_i,y_i) \,,
\end{align} 
where $\q^4 = \frac{1}{12} \q_{a}^\alpha \q_\a^b \q_b^\b \q_\b^a$ like in \re{Qdef}. This is of course not the only solution of the Ward identities \p{I-inv}. The right-hand side of \p{calA} could also contain terms involving the anti-chiral variables $\bq_i$. However,  we will soon show that the particular solution \p{calA} of \p{I-inv} also satisfies  the other half of the Ward identities \p{I-inv1} and thus provides the  complete and unique $\cI_4$.
The explicit form of the function $A_4(x_i,y_i)$ 
can be found by matching \p{calG4-loop}, \p{calI} and \p{calA} with the lowest component \p{G4-loop}  (see Sect.~\ref{s2.4} for more details). 

Let us demonstrate that  $\cI_4$ defined in \p{calI}  with $\cA_4$ given by \p{calA} is also annihilated by the other half of the odd generators, Eq.~\p{I-inv1}, provided that the function $\cA_4$ is invariant under the bosonic part of the superalgebra, the conformal and $R-$symmetry.  To show this, we hit \p{calI} with $\bar Q$ or $S$, whose explicit expressions can be obtained by conjugating $Q$ and $\bar S$ in \p{generators}.
Once these generators have gone through  the differential operator in \p{calI} and have reached $\cA_4$, they annihilate it. Indeed,    $\bar Q$ and $S$ are given by a linear combination of terms proportional to $\pa/\pa \bq_i$, $\q_i$ or $\q_i\q_i \pa/\pa \q_i$, each of which 
gives zero on the right-hand side of \p{calA}. Further, as we show in Appendix \ref{algebra}, the commutator of $\bar Q$ or $S$ with the differential operator in \p{calI} annihilates $\cA_4$ provided that it is invariant under conformal and $R-$symmetry transformations. We can use
the explicit form of $A_4$ (see  Eq.~\p{232} below) to verify that this is indeed the case.  Finally, evoking once more the uniqueness of the supersymmetric extension of the bosonic four-point correlation function, we can claim that it is given by \p{calI}.

Clearly, we could have constructed the supersymmetric extension using the generators  $\bar Q$ and $S$ instead of $Q$ and $\bar S$  since the defining relations \p{I-inv} and \p{I-inv1} are obviously symmetric under the exchange of these generators. Thus, the invariant \p{calI} should admit another representation 
\begin{align}\label{241}
 \mathcal I_4\left(x_i,\q_i,\bq_i,y_i\right) =\bar Q^4 \bar Q'{}^4 S^4 S'{}^4 \big[  \bq_1^4 \, \bq_2^4 \, \bq_3^4 \, \bq_4^4 A_4(x_i,y_i)\big],
\end{align}
where $A_4(x_i,y_i)$ is the same function as in \p{calA}.  At first sight, it is not obvious why the two expressions \p{calI} and \p{241} are equivalent. The reason is that both of them are extensions of the same four-point bosonic correlator, but we know that this extension is unique. In this way we have proven the rather non-trivial identity
\begin{align}\label{242}
Q^4 Q'{}^4 \bar S^4 \bar S'{}^4  \big[  \q_1^4 \, \q_2^4 \, \q_3^4 \, \q_4^4 A_4(x_i,y_i)  \big]=\bar Q^4 \bar Q'{}^4 S^4 S'{}^4 \big[  \bq_1^4 \, \bq_2^4 \, \bq_3^4 \, \bq_4^4 A_4(x_i,y_i)\big]  \,.
\end{align}
Expanding these two forms in the odd variables results in two, superficially different but equivalent forms of the components of the supercorrelator (see an example in \p{JOOO1} 
and \p{J'} below).

We remark that the energy-momentum supermultiplet $\mathcal T(x,\q,\bq,y)$ is real in the sense of the combined complex and harmonic conjugation in harmonic superspace (see \cite{hh,Howe:1995md}), and so is  the invariant $\mathcal I_4$. Assuming the reality of the function $A_4(x_i,y_i)$, we see an additional reason why the two forms of  $\mathcal I_4$ must be equivalent.

\subsection{Matching condition}   \label{s2.4}

Combining together  \re{calG4-loop}, \re{calI} and \re{calA}, we conclude that the four-point correlation function $\mathcal G_4^{\nr}$ is 
determined by the scalar function $A_4(x_i,y_i)$ depending only on the bosonic variables.
To identify   $A_4(x_i,y_i)$, we compare the lowest component of $\mathcal G_4^{\nr}$, corresponding 
to $\theta_i=\bar\theta_i=0$, with the four-point correlation function $G_4^{\nr}$ of half-BPS operators, Eq.~\re{match'}:
\begin{align}\label{match}
Q^4 Q'{}^4 \bar S^4 \bar S'{}^4
\big[  \q_1^4  \q_2^4 \q_3^4 \q_4^4 A_4(x_i,y_i)  \big]\big|_{\theta_i=\bar\theta_i=0} = {\Phi(u,v) \over uv} (\zeta-w)(\zeta-\bar w)(\bar \zeta-w)(\bar \zeta-\bar w)\,.
\end{align}
To obtain a non-vanishing result on the left-hand side, the 16 generators  $Q$ and $\bar S$ must remove all 16 $\q$'s  in  the square brackets. Therefore,  only the terms with  $\pa_\q$ in the generators \re{generators} contribute. In this way we find for the left-hand side  of \re{match}
\footnote{This result can easily be obtained by fixing a conformal and $SU(4)$ gauge in which the coordinates of 3 points are chosen as $x_1 =y_1=1$, $x_2 = y_2=\infty$, $x_3=y_3= 0$, and the 
coordinates  of the fourth point, $x_{4}^{\da\a}$ and $y_{4}^{a'a}$  are diagonal matrices with $\zeta, \bar \zeta$ and $w,\bar w$ on their diagonals, respectively. 
The details of the calculation can be found in Appendix~B of  Ref.~\cite{Eden:2011we}. }
\begin{align}\notag
 \big(\sum \partial_{\theta_i}\big)^4   \big(\sum x_i\partial_{\theta_i}\big)^4  & \big(\sum \partial_{\theta_i} y_i\big)^4 \big(\sum x_i\partial_{\theta_i}  y_i\big)^4 \big[  \q_1^4  \q_2^4 \q_3^4 \q_4^4 A_4(x_i,y_i)  \big]\big|_{\theta_i=\bar\theta_i=0} 
 \\[2mm] 
 = {} &  A_4(x_i,y_i)     (x_{13}^2 x_{24}^2 y_{13}^2 y_{24}^2)^2 (\zeta-w)(\zeta-\bar w)(\bar \zeta-w)(\bar \zeta-\bar w)\,.  \label{244}
\end{align}
Together with \re{match}, this implies 
\begin{align}\label{232}
A_4(x_i,y_i)  = {\Phi(u,v) \over uv} {1\over ( x_{13}^2 x_{24}^2 y_{13}^2 y_{24}^2)^2}\,.
\end{align}
The substitution of \p{232} into  \re{calG4-loop}, \re{calI} and \re{calA} yields the following result for the anomalous part of the  four-point correlation function \footnote{Notice that the rational part \p{calG-Born} admits an analogous representation.
After pulling out a propagator factor from $\mathcal G_4^{\rt}$ as in \p{calG4-loop}, we obtain the
invariant $\mathcal I_4$ as a polynomial in two variables, $U/u$ and $V/v$ (see \p{210}). Then, the relation  \p{calI}
fixes the corresponding function $A_4$ to be this polynomial divided by $(\zeta-w)(\zeta-\bar w)(\bar \zeta-w)(\bar \zeta-\bar w)$. The uniqueness of the supersymmetric extension guarantees that the result will be the same as in \p{calG-Born}. } 
\begin{align}\label{G4-main}
\mathcal G_4^{\nr} =  \lr{\frac{y^2_{13}}{\hat x^2_{13}}\frac{y^2_{24}}{\hat x^2_{24}}}^2 Q^4 Q'{}^4 \bar S^4 \bar S'{}^4
\left[    {\q_1^4  \q_2^4 \q_3^4 \q_4^4 \over ( x_{13}^2 x_{24}^2 y_{13}^2 y_{24}^2)^2}{\Phi(u,v) \over uv}  \right],
\end{align}
where $Q$ and $\bar S$ are the differential operators defined in \re{generators} and  the function $\Phi(u,v)$ was 
introduced in \re{G4-loop}. It is straightforward to verify that the relations \p{cross} ensure the crossing symmetry of $\mathcal G_4^{\nr}$.

We remark that the expression in the square brackets in \p{G4-main}, i.e., the function $\cA_4$ from \p{calI}, is invariant under both the conformal and R symmetry groups. Indeed, $\Phi(u,v)/(uv)$ is conformally invariant, while the denominator in the first factor compensates the conformal and $SU(4)$ weights of the nilpotent numerator. As explained in Section~\ref{s26}, this property is essential for proving that \p{calI} provides the complete solution of the Ward identities \p{I-inv} and \p{I-inv1}.

We would like to make an important comment about the form of the right-hand side of \p{match}. It originates from the lowest component \p{match'} of the anomalous 
part of the correlator. We can now invert the logic and claim that the specific form of \p{match'}  is a corollary of the procedure of supersymmetrization of the bosonic 
correlator. Indeed, let us assume that the right-hand side of \p{match'} has the following generic form $P(U,V)=\sum_{0\leq m+n\leq2}\Phi_{mn}(u,v) U^m V^n$ in 
terms of the harmonic cross-ratios defined in \p{210}. The degree of this polynomial in $U,V$ cannot exceed 2 because otherwise the propagator prefactor in \p{match'} 
will not be able to cancel the $y-$singularities.\footnote{We recall that the $y$'s are auxiliary variables on a coset of the $R-$symmetry group $SU(4)$. They carry the 
various irreducible representations appearing in the tensor product $\mathbf{20'}\times \mathbf{20'}$. Hence, the dependence on $y$ must be polynomial.} Then, 
matching $P(U,V)$ with \p{244} will result in a {\it singular} function $A_4(x,y)$. This will lead to unacceptable $y-$singularities in the higher components in the expansion 
of the supercorrelator. In order to avoid this, we conclude that $P(U,V)$ must have the factorized form of the right-hand side of \p{match}. This is another way of proving 
the so-called `partial non-renormalization theorem' of  Ref.~\cite{Eden:2000bk} (see also \cite{Dolan:2004mu}  for a similar argument). 

As explained in Section~\ref{s24}, what we have obtained here is the {\it unique} supersymmetric completion of the bosonic correlation function \p{Scalar4ptCorr}. Its explicit expression \re{G4-main} is one of the main results of this paper. In the next section, we examine  the detailed properties of some of its components.
 
\section{Extracting four-point correlation functions}

Despite the compact form of  the correlation function \re{G4-main}, extracting its various components is a nontrivial technical task. However, it becomes considerably simpler for the correlation function
\begin{align}\label{hatG4}
\widehat{\mathcal G}_4 
=  \mathcal G_4\big|_{\theta_{3,4}=\bar\theta_{3,4}=0}= \vev{\mathcal T(1) \,\mathcal T(2)\, O(3)\, O(4)}  \,.
\end{align}
It depends on the bosonic coordinates $x_i$ and $y_i$  (with $i=1,\dots,4$) and the Grassmann variables  $\theta_{1,2}$ and $\bar\theta_{1,2}$. Moreover, this particular class of correlation functions has been used in \cite{Belitsky:2013xxa,Belitsky:2013bja,Belitsky:2013ofa} for computing event shape functions (see also Sect.~\ref{s4} below for an alternative treatment).

The correlation function \re{hatG4} has specific transformation properties under the conformal  and $R-$symmetry groups.
We can exploit them by first computing $\widehat{\mathcal G}_4$ for some special configuration of $x_i$ and $y_i$ and then restoring
its general covariant form.
 
\subsection{Gauge fixing}\label{sect:gauge}

As follows from \re{CSUSY}, the coordinates of the scalar operators $O(3)$ and $O(4)$ in \re{hatG4} do not vary under the superconformal
transformations \re{CSUSY},  $\delta x_{3,4}=\delta y_{3,4}=0$  for $\theta_{3,4}=\bar\theta_{3,4}=0$.  This allows us
to fix the conformal and $SU(4)$ gauge
\begin{align}\label{gauge}
x_{3,\alpha\dot\alpha} = y_{3,aa'}=0 \,,\qquad\qquad x_{4,\alpha\dot\alpha} , \ y_{4,aa'} \to \infty\,.
\end{align}
Then, the general expression for the anomalous correlation function \re{G4-main} simplifies to
\begin{align}\label{G4-gauge}
\widehat{\mathcal G}_4^{\nr}  = \lr{\frac{y^2_{1}}{\hat x^2_{1}}\frac{y^2_{4}}{x^2_{4}}}^2 Q^4Q'{}^4 \bar S^4\bar S'{}^4
\left[    {\q_1^4  \q_2^4 \q_3^4 \q_4^4 \over (x_{1}^2 x_{4}^2 y_{1}^2 y_{4}^2)^2}{\Phi(u,v) \over uv}  \right]\bigg|_{\theta_{3,4}=\bar\theta_{3,4}=0},
\end{align}
where $\hat x_{1}^{\alpha\dot\alpha} = x_{1}^{\alpha\dot\alpha}  - \q^{a\alpha}_{1} (y^{-1}_{1})_{aa'} \bq^{a'\dot\alpha}_{1}$ and the conformal cross-ratios take the form, $u=x_{12}^2/x_1^2$ and $v=x_2^2/x_1^2$.  Here we 
took into account that, since for $\theta_{3,4}=\bar\theta_{3,4}=0$ the generators of superconformal transformations \re{generators} do not contain 
derivatives with respect to $x_3$ and $y_3$, the gauge \re{gauge} can be imposed inside the square brackets in \re{G4-gauge}.
In addition,   we can simplify the expression for the generator  $\bar S'$ in \re{generators} as
\begin{align}
\bar S_{b'\dot\beta} = x_{4,\alpha\dot\beta} y_{4,b'}{}^a{\partial \over \partial \theta_{4,\alpha}^a} + \dots \,,\qquad   \qquad
\bar S'{}^4 = (x_4^2 y_4^2)^2 (\partial_{\theta_{4}})^4 + \dots\,,
\end{align}
where the dots denote terms subleading for $x_4,y_4\to\infty$. Note that this limit eliminates the non-linear terms $\bq\bq \pa_{\bq}$ in $\bar S'$, which would otherwise complicate the Grassmann expansion in \p{G4-gauge}.  We also observe that for $x_3=y_3=0$, all generators in \re{G4-gauge} except
$Q$ do not involve derivatives with respect to $\theta_3$. Therefore, evaluating \re{G4-gauge} we are allowed to replace
$Q^4\to (\partial_{\theta_{3}})^4$. 

In this way, we obtain from \re{G4-gauge}
\begin{align}\label{f} 
&  \widehat{\mathcal G}_4^{\nr}  =  \lr{y^2_{4} \over x^2_{4}}^2  Q'{}^4 \bar S^4 
\left[ \q_1^4  \,\q_2^4 \,f(x_1,x_2)  \right] ,  && 
 f(x_1,x_2) =  {\Phi\lr{{x_{12}^2/x_1^2}\,, {x_{2}^2/x_1^2}} \over  (x_1^2)^2 x_{12}^2 x_2^2 }
\end{align}
where 
we used the relations 
$ Q'{}^4 \bar S^4(y_1^2/\hat x_1^2 )=0$ and $\q_1^4/\hat x_1^2 =\q_1^4/x_1^2$. Here  the generators $Q'$ and $\bar S$ only act at points $1$ and $2$ and are given 
by the following simplified expressions
 \begin{align}\label{QS} 
&Q_\a^{a'} = \sum_{i=1,2}\left( y^{a'a}_i \frac{\partial}{\partial \theta^{\alpha a}_i}+
\frac{\partial}{\partial x^{\da\a}_i}\bq^{\da a'}_i\right),\qquad \qquad
\bar S_a^{\da} = \sum_{i=1,2}\left( x^{\da\a}_i\frac{\partial}{\partial \q^{\a a}_i} -\bq^{\da a'}_i \frac{\partial}{\partial y^{a'a}_i} \right).
\end{align}
Notice that the function $f(x_1,x_2)$ defined in \re{f} does not depend on the $y-$coordinates and satisfies 
$f(x_1,x_2)=f(x_2,x_1)$ in virtue of \re{cross}.

 \subsection{Single current insertion}
 
 In this subsection, we apply the general formula \re{f}  to compute the correlation functions involving three half-BPS operators at points 2,3,4 
 and an $R-$symmetry current or an energy-momentum tensor at point 1. 
  
\subsubsection{Single R-current insertion}

As the first non-trivial example, let us apply \re{f} to obtain the anomalous part of the correlation function \re{JOOO} involving a single
insertion of the $R-$current
\begin{align}\notag\label{CorrjDiff}
G_{\alpha\dot\alpha,aa'} & = \vev{ J_{\alpha\dot\alpha,aa'}(1)\, O(2)\, O(3)\, O(4)}^{\nr} 
\\[2mm]&
= \lr{y^2_{4} \over x^2_{4}}^{\!\!2}   
\left[ 
(\partial_{\bar\theta_1})_{\dot\alpha a'}(\partial_{\theta_1})_{\alpha a} + \frac12 (\partial_{x_1})_{\alpha\dot\alpha} (\partial_{y_1})_{a'a}
\right]  
Q'{}^4 \bar S^4 
\left[ \q_1^4  \q_2^4 \,f(x_1,x_2)  \right] \big|_{\theta_{1,2}=\bar\theta_{1,2}=0} \,.
\end{align}
We recall that the second term inside the square brackets guarantees the current conservation,
$(\partial_{x_1})^{\alpha\dot\alpha} G_{\alpha\dot\alpha,aa'} =0$. To evaluate (\ref{CorrjDiff}) we begin with the identity
\begin{align} \label{Id1}
\frac{\partial}{\partial \bar\theta_1^{\da a'}} Q'{}^4 \bar S^4 
\left[ \q_1^4  \q_2^4 \,f(x_1,x_2)  \right]  \bigg|_{\bar\theta_{1,2}=0}  
 &= -
   \frac{\partial}{\partial x_1^{\b\da}} (Q'{}^3)_{a'}^\b \bar S^4  \left[ \q_1^4  \q_2^4 \,f(x_1,x_2)  \right] 
 \bigg|_{\bar\theta_{1,2}=0}\, ,
\end{align}
{where $(Q'{}^3)_{a'}^\b = \frac{1}{3} Q^\b_{b'} Q^{b' \g} Q_{a' \g}$ with} 
$Q'$ and $\bar S$ defined in (\ref{QS}). Here we took into account that $ \{\partial_{\bar\theta_1^{\da a'}}, \bar S_a^{\db}\} = -\delta_{\da}^{\db}\partial_{y_1^{a'a}}$ does not contribute since the expression inside the square brackets in \p{Id1} does not
depend on $y_1$. In the same way, we find
\begin{align} \label{Id2}
\frac{\partial}{\partial y_1^{a' a}} Q'{}^4 \bar S^4  \left[ \q_1^4  \q_2^4 \,f(x_1,x_2)  \right]  \bigg|_{\bar\theta_{1,2}=0}  
 &= -
  \frac{\partial}{\partial \q_1^{\b a}} (Q'{}^3)_{a'}^\b \bar S^4 \left[ \q_1^4  \q_2^4 \,f(x_1,x_2)  \right] \bigg|_{\bar\theta_{1,2}=0}\,.
\end{align}
Combining the last two relations we obtain
\begin{align}\label{jooo}
G_{\alpha\dot\alpha,aa'} =   
\left(\frac{\partial}{\partial \theta_1^{\a a}}\frac{\partial}{\partial x_1^{\b\da}}-\frac{1}{2} \frac{\partial}{\partial \q_1^{\b a}}\frac{\partial}{\partial x_1^{\a\da}} \right)  
(Q'{}^3)_{a'}^\b \bar S^4  \bigg[ \q_1^4  \q_2^4 \,  f(x_1,x_2) \lr{\frac{y^2_{4}}{x^2_{4}}}^{\! \!2} \,\bigg]
 \bigg|_{\bar\theta_{1,2}=0}\,.
\end{align}
As a quick check, we apply $(\partial_{x_1})^{\alpha\dot\alpha} $ to both sides of this equation and verify that $(\partial_{x_1})^{\dot\alpha\alpha} G_{\alpha\dot\alpha,aa'} =0$, as expected due to the $R-$current conservation. 

Remarkably, relation \re{jooo}
admits an  equivalent representation in which this property becomes manifest. We observe that for $\bar\theta_{1,2}=0$
the generators \re{QS} are reduced to differential operators acting on the Grassmann variables. This 
allows us to write the correlation function \re{jooo}  as a total derivative with respect to $x_1$
\begin{align}\label{G-M}
G_{\alpha\dot\alpha,aa'} = (\partial_{x_1})_{\da}^\b
\bigg[\cM_{\a\b, aa'}  f(x_1, x_2)  \lr{\frac{y^2_{4}}{x^2_{4}}}^{\! \!2}
 \bigg],
\end{align}
where we introduced the  tensor
\begin{align}\label{MatrixJdiff}\notag
\cM_{\a\b, aa'} & =
 \left(\epsilon_{\g\b}\frac{\partial}{\partial \theta_1^{\a a}} -\frac12\epsilon_{\a\b}\frac{\partial}{\partial \q_1^{\g a}}  \right)
(Q{'}{}^3)_{a'}^\g \bar S^4 \left(
  \q_1^4 \q_2^4 \right)\bigg|_{\bar\theta_{1,2}=0} 
\\  
 & =
 (\partial_{\q_1})_{(\a a} (Q{'}{}^3)_{\b)a'} \bar S^4 \left(
  \q_1^4 \q_2^4 \right)\bigg|_{\bar\theta_{1,2}=0}
  \,. 
\end{align}
Since the expression in the second line involves 8 Grassmann derivatives acting on a polynomial of Grassmann degree 8,
$\cM_{\a\b, aa'} $ does not depend on  $\theta$ anymore. Most importantly, the tensor 
\re{MatrixJdiff} is traceless with respect to its Lorentz indices
\begin{align}
 \epsilon^{\a\b} \cM_{\a\b, aa'}  = 0\,.
\end{align}
We can use \re{G-M} together with the identity $(\partial_{x_1})^{\dot\alpha\alpha}(\partial_{x_1})_{\dot\alpha}^{\beta}=-\Box_{x_1}\epsilon^{\alpha\beta}$ to verify that this property makes the current conservation manifest,
$(\partial_{x_1})^{\dot\alpha \alpha} G_{\alpha\dot\alpha,aa'} =0$. The evaluation of \re{MatrixJdiff} gives 
\begin{align}
 \cM_{\a\b,aa'} = (x_{12} x_1)_{(\a\b)}\left[ (y_1)_{aa'}( y_{12}^2 x_2^2-x_{12}^2y_2^2) + (y_{12})_{aa'}(x_1^2 y_2^2-y_1^2 x_2^2)\right],
\end{align}
where $(\alpha\beta)$ denotes weighted symmetrization. Substituting this relation into \re{G-M} we obtain the anomalous part of the correlation function \re{JOOO} in the gauge \re{gauge}.

The final step is to restore the dependence on the points $3$ and $4$. This can be done
by performing a conformal transformation in \re{G-M} combined with a  $SU(4)$ rotation.\footnote{Equivalently, we might 
write the general expression for the correlation function in term of the basis tensors $X$ and $Y$ from \p{XY},  consistent with the conformal properties of the operators and fix the free parameters by matching it with \re{G-M} in the gauge \re{gauge}.} We finally arrive at the following expression
\begin{align}\notag\label{JOOO1}
& \vev{ J_{\alpha\dot\alpha,aa'}(1)\, O(2)\, O(3)\, O(4)}^{\nr} 
\\[2mm]
&  = {1 \over 4 } (\px{1})_{\da}^\b \left\{ \left(  y^2_{23} y^2_{34} Y_{124} - u\,  y^2_{23}  y^2_{24} Y_{134} - v\,  y^2_{24} y^2_{34} Y_{123}\right)_{aa'} [X_{124},X_{134}]_{(\a\b)}  \frac{\Phi(u,v)}{x^2_{23} x^2_{24} x^2_{34}} \right\}\,,
\end{align}
where the structures $X_{ijk}$ and $Y_{ijk}$ were defined in \p{XY}.

In Section~\ref{s26}, we pointed out the existence of the equivalent form \p{241} of the supercorrelator. If we use it instead of \p{G4-main}, the expression for the single current insertion will differ from \p{JOOO1} by the exchange of chiral and antichiral indices,
\begin{align}\notag\label{J'}
& \vev{ J_{\alpha\dot\alpha,aa'}(1)\, O(2)\, O(3)\, O(4)}^{\nr} 
\\[2mm]
&  = {1 \over 4 } (\px{1})_{\a}^\db \left\{ \left(  y^2_{23} y^2_{34} Y_{124} - u\,  y^2_{23}  y^2_{24} Y_{134} - v\,  y^2_{24} y^2_{34} Y_{123}\right)_{aa'} [X_{124},X_{134}]_{(\da\db)}  \frac{\Phi(u,v)}{x^2_{23} x^2_{24} x^2_{34}} \right\}\,.
\end{align}
It is not immediately obvious but nevertheless true that the two expressions coincide. This is a manifestation of the general identity \p{242}.


\subsubsection{Single energy-momentum tensor insertion}

Let us now examine the correlation function involving a single energy-momentum tensor insertion
\begin{align}
G_{\alpha\beta,\dot\alpha\dot\beta} =\vev{T_{\alpha\beta,\dot\alpha\dot\beta}(1)\,O(2)\, O(3)\, O(4)}^{\nr} \,.
\end{align}
We recall that the energy-momentum tensor appears as a particular component in the expansion of the superfield $\cT$ and its correlation function 
can be extracted with the help of the differential operator \re{diffT}. This leads to 
\begin{align}\notag\label{G-T}
  G_{\alpha\dot{\alpha},\beta\dot{\beta}} = \left[  - (\partial_{\q_1})_{\a}^{a}   (\partial_{\q_1})_{ \b a} (\partial_{\bq_1})_{\da a'}
 (\partial_{\bq_1})_{\db}^{a'}  
-
  (\partial_{\q_1})_{(\a}^{a} (\partial_{x_1})_{\b)(\db}(\partial_{y_1})_{aa'} (\partial_{\bq_1})_{\da)}^{a'} \right. {}&
\\ 
   \left.  + \frac1{6}  (\partial_{x_1}) _{(\a\da} (\partial_{x_1}) _{\b)\db}
 (\partial_{y_1})_{aa'} (\partial_{y_1})^{a'a}\right] {}& \widehat{\mathcal G}_4^{\nr}\big|_{\theta_{1,2}=\bar\theta_{1,2}=0}\,.
\end{align}
Here the last two terms inside the square brackets subtract the contribution of the descendants given by  total derivatives 
of lower components of the energy-momentum  supermultiplet $\cT$. They are required to ensure the current conservation
$(\partial_{x_1})^{\alpha\dot\alpha} G_{\alpha\dot{\alpha},\beta\dot{\beta}} =0$.

As in the case of a single $R-$current insertion, in order to evaluate \re{G-T} we fix the gauge \re{gauge} and replace $\widehat{\mathcal G}_4^{\nr}$ in \re{G-T} 
by its expression \re{f}. Making use of the identities \re{Id1} and \re{Id2},  after some algebra we find
\begin{align}\label{G-T-gauge}
G_{\alpha\dot{\alpha},\beta\dot{\beta}} =   (\partial_{x_1})_{\da}^{\delta}(\partial_{x_1})_{\db}^{\g}\bigg[\cM_{\a\b \delta\g}  f(x_1, x_2)  \lr{\frac{y^2_{4}}{x^2_{4}}}^{\! \!2}
 \bigg]\,,
\end{align}
with the tensor $\cM_{\a\b}^{\delta\g}={\cM_{\a\b \sigma\tau}\epsilon^{\delta\sigma}\epsilon^{\g\tau}}$ given by
\begin{align}\notag
\cM_{\a\b}^{\delta\g} = 
\big[-(\partial_{\q_1})_\a^a(\partial_{\q_1})_{\b a} (Q {'}{}^2)^{\delta\g}
+\delta^{(\delta }_{(\a} (\partial_{\q_1})_{\b)}^a(\partial_{\q_1})_{\g' a}(Q{'}{}^2)^{\g)\g'} &
\\
-\ft16 \delta_{(\a}^{\delta} \delta_{\b)}^{\g} (\partial_{\q_1})_{\delta'}^a(\partial_{\q_1})_{\g' a}(Q{'}{}^2)^{\delta'\g'}\big] & \bar S^4 \left(
  \q_1^4 \q_2^4 \right)\bigg|_{\bar\theta_{1,2}=0}\,,
\end{align}
{with $(Q'{}^2)^{\a\b} = Q^\a_{a'} Q^{a'\b}$}.
Going through a lengthy calculation we arrive at the surprisingly simple result
\begin{align}
\cM_{\a\b\delta\g}  = 
-\frac13  (x_1x_2)_{((\a\delta} (x_1x_2)_{\b\g))}  y_2^2\,,
\end{align}
where the double-parentheses notation indicates symmetrization of all four indices, so that 
\begin{align}
\epsilon^{\a\b}  \cM_{\a\b\delta\g}=0\,.
\end{align}
This property implies that the correlation function \re{G-T-gauge} satisfies the relations
\begin{align}
 G_{\alpha\dot{\alpha},\beta\dot{\beta}} -  G_{\beta\dot{\beta},\alpha\dot{\alpha}}= \epsilon^{\a\b}\epsilon^{\da\db} G_{\alpha\dot{\alpha},\beta\dot{\beta}} =(\partial_{x_1})^{\dot\alpha \alpha} G_{\alpha\dot{\alpha},\beta\dot{\beta}} = 0\,,
\end{align}
which ensure that the energy-momentum tensor is symmetric, traceless and conserved.

Relation \re{G-T-gauge} has been obtained in the conformal gauge \re{gauge}. The covariant expression for this correlation function is
\begin{align}\notag \label{T1-con}
& \vev{T_{\alpha\beta,\dot\alpha\dot\beta}(1)\, O(2)\, O(3)\, O(4)}^{\nr} 
\\[2mm] & \qquad ={1 \over 4 }
 (\partial_{x_1})^{\g}_{\da}(\partial_{x_1})^\delta_{\db}  \left[ [X_{134},  X_{124}]_{((\a\b}  [X_{134},  X_{124}]_{\g\delta))}   \Phi(u,v)\frac{x_{12}^2x_{14}^2}{x_{24}^2} \frac{y_{23}^2 y_{34}^2 y_{42}^2}{x_{23}^2 x_{34}^2 x_{42}^2} \right].
\end{align}
Indeed, it is straightforward to verify that it has the correct properties under conformal and $SU(4)$ transformations and
coincides with \re{G-T-gauge} in the gauge \re{gauge}.
 
\subsubsection{Analogy with conformal field equations}

{The particular form of the correlation functions \p{G-M} and \p{G-T-gauge} suggests the following analogy with the conformally covariant Maxwell and Weyl equations 
\begin{align}\label{3.24}
(\pa_x)^\b_\da F_{\a\b} = J_{\a\da}\,, \qqqquad   (\pa_x)^\gamma_\da (\pa_x)^\delta_\db C_{\a\b\gamma\delta} = T_{\a\da,\, \b\db}\,.
\end{align}
Here $F_{\a\b}$ and $C_{\a\b\gamma\delta}$ are the self-dual  (or chiral) Maxwell and Weyl tensors, respectively, fully symmetrized in their spinor indices. This property immediately implies the conservation of the corresponding sources $J_{\a\da}$ and $T_{\a\da,\, \b\db}$. Further, the fact that the tensors $F_{\a\b}$ and $C_{\a\b\gamma\delta}$ belong to the representations $(1,0)$ and $(2,0)$ of the Lorentz group, respectively, together with the appropriate conformal transformation laws,  assures the covariance of the field equations \p{3.24}.\footnote{A similar observation for three-point functions with a current or an energy-momentum tensor is made in \cite{Erdmenger:1997wy}.}

Another remark concerns the purely chiral form of the matrices $\cM$ entering \p{G-M} and \p{G-T-gauge}. In the case of the field equations \p{3.24} there exists an 
equivalent anti-chiral form, e.g., $(\pa_x)^\db_\a \tilde F_{\da\db} = J_{\a\da}$ obtained by complex conjugation (assuming that the current $J_{\a\da}$ is hermitean). 
Similarly, had we preferred to build the superconformal invariants \p{calI}
 with the generators $\bar Q$ and $S$ instead of $Q$ and $\bar S$, we would have obtained Eqs.~\p{G-M} and \p{G-T-gauge} with anti-chiral matrices 
 ${\ }\overline{\!\!\cM}$ (see \p{J'} for an example). The equivalence of the two forms relies on some non-trivial identities.  
}

Comparing \re{JOOO1} and \re{T1-con} we observe that the two correlation functions have a similar form. This 
suggests that the correlation function with a single insertion of a conserved current of spin $s$ possesses the universal form  
\begin{align}\label{G-simple}
 \vev{J_{\a_1\da_1\dots\a_s\da_s}(1)\, O(2)\, O(3)\, O(4)} =
 (\partial_{x_1})^{\b_1}_{\da_1}\dots (\partial_{x_1})^{\db_s}_{\da_s}  \big[ \cM_{\a_1\b_1\dots \a_s\b_s}\Phi(u,v)]\,,
\end{align}
where the tensor $\cM_{\a_1\b_1\dots \a_s\b_s}$ is completely symmetric with respect to all of its indices and, most importantly,
it is independent of the dynamical details of the  theory. We will show in the next subsection that a similar relation also holds for
correlation functions with double insertion of the conserved currents.
  
\subsection{Double current insertion}

To simplify the analysis, in what follows we impose the gauge condition \re{gauge}. In close analogy with
\re{CorrjDiff}, the correlation function involving two insertions of the $R-$current  can be obtained from \re{f} by applying
the same differential operator at points $1$ and $2$
\begin{align}\notag\label{JJOO}
G_{JJOO} 
& = \vev{ J_{\a_1\da_1,a_1a_1'}(1)\, J_{\a_2\da_2,a_2a_2'}(2)\, O(3)\, O(4)}^{\nr}
\\[2mm]
 &=\prod_{i=1,2}\left[ {(\partial_{\theta_i})_{\a_i a_i}} {(\partial_{\bar\theta_i})_{\da_i a_i'}}
+
 \ft12 {(\partial_{x_i})_{\a_i\da_i}} {(\partial_{y_i})_{a_i' a_i}}\right]\widehat{\mathcal G}_4^{\nr}\big|_{\theta_{1,2}=\bar\theta_{1,2}=0} \,.
\end{align}
Replacing $\widehat{\mathcal G}_4^{\nr}$ with its expression \re{f}, we evaluate the derivatives with respect to $\bar\theta_i$
and $y_i$ with the help of the identities \re{Id1} and \re{Id2} to obtain
\begin{align}\notag
G_{JJOO} 
& = \prod_{i=1,2}\left[ {(\partial_{\theta_i})_{\a_i a_i}} {(\partial_{ x_i})_{\b_i\da_i}}-\ft12  {(\partial_{\q_i})_{\b_i a_i}} {(\partial_{x_i})_{\a_i\da_i}} \right]Q_{a'_2}^{\b_1} Q_{a'_1}^{\b_2} \bar S^4\left[
  \q_1^4\, \q_2^4 \, f(x_1, x_2)  \lr{\frac{y^2_{4}}{ x^2_{4}}}^{\! \!2}
 \right]\bigg|_{\theta_{1,2}=\bar\theta_{1,2}=0}\,.
\end{align}
This relation can be rewritten in a very suggestive form involving derivatives with respect to $x_{1,2}$
\begin{align}\label{G-2J}
G_{JJOO}  
& =(\partial_{x_1})_{\da_1}^{\b_1}(\partial_{x_2})_{\da_2}^{\b_2}
\bigg[\cM_{\a_1\b_1\a_2\b_2, a_1a_1'a_2a_2'}  f(x_1, x_2)  \lr{\frac{y^2_{4}}{x^2_{4}}}^{\! \!2}
 \bigg],
\end{align}
with the $\cM-$tensor defined as
\begin{align}
 \cM_{\a_1\b_1\a_2\b_2, a_1a_1'a_2a_2'} = - (\partial_{\theta_1})_{(\a_1 a_1}  Q_{\b_1)a'_2} 
 (\partial_{\theta_2})_{(\a_2 a_2} Q_{\b_2)a'_1}  \bar S^4\left(
  \q_1^4 \q_2^4 \right)\bigg|_{\bar\theta_{1,2}=0}\,.
\end{align} 
The very fact that this tensor is symmetric with respect to two pairs of indices, $\a_1,\b_1$ and $\a_2,\b_2$, ensures 
that the correlation function \re{JJOO} vanishes under the action of operators $(\partial_{x_1})^{\alpha_1\dot\alpha_1}$ 
and $(\partial_{x_2})^{\alpha_2\dot\alpha_2}$, as implied by current conservation. Explicitly, 
we find
\begin{align}\notag
\cM_{\a_1\b_1\a_2\b_2, a_1a_1'a_2a_2'} &=(y_1\cdot y_2)\epsilon_{a_1'a_2'}\epsilon_{a_1a_2}\big[ (x_2x_1)_{\a_1\b_1}(x_1x_2)_{\a_2\b_2}-\epsilon_{\a_1\a_2}\epsilon_{\b_1\b_2}
x_1^2 x_2^2 \big]
\\\notag
&
+ (y_1)_{a_1a_2'} (y_1)_{a_2 a_1'} \epsilon_{\a_1\a_2} (x_1 x_2)_{\b_1\b_2} x_2^2+ (y_2)_{a_1a_2'} (y_2)_{a_2a_1'} (x_2 x_1)_{\a_1\a_2} \epsilon_{\b_1\b_2} x_1^2 
\\
&
-2 (y_1)_{a_1a_2'} (y_2)_{a_2a_1'}  \epsilon_{\a_1\a_2}\epsilon_{\b_1\b_2}
(x_1\cdot x_2)^2
 \,,  \label{3.29}
\end{align}
where the symmetrization with respect to the Lorentz indices  $\a_1,\b_1$ and $\a_2,\b_2$ is tacitly assumed.
To save space, we do not present the covariant form of \re{3.29} here. 

This analysis can be extended to correlation functions  involving the energy-momentum  tensor, 
$\vev{J(1)\, T(2)\, O(3)\, O(4)}$ and  $\vev{T(1)\, T(2)\, O(3)\, O(4)}$.
For our purposes, it is important that these correlation functions (as well as \re{G-2J}) admit the following representation
\begin{align}\label{JJOO-gen}
& \vev{J_{\a_1\da_1\dots\a_s\da_s}(1)\, J_{\b_1\db_1\dots\b_{s'}\db_{s'}}(2)\, O(3) O(4)}^{\nr} =
\prod_{i=1}^s (\partial_{x_1})^{\g_i}_{\da_i} \prod_{k=1}^{s'} (\partial_{x_2})^{\delta_k}_{\db_k} \big[ \cM_{\{\a\}\{\b\}\{\g\}\{\delta\} }\Phi(u,v)]\,.
\end{align} 
In other words, they are given by total derivatives with respect to the current insertion points. Here, all information about
the choice of the currents is encoded in the tensor $\cM$ that only carries chiral indices $\{\alpha\},\{\beta\},\{\gamma\},\{\delta\}$.
Current conservation translates into the invariance of $\cM$ 
under the exchange of any pair of indices belonging to $\{\a\}\cup \{\g\}$ and  $\{\b\}\cup \{\delta\}$.  
As already mentioned, there exists an equivalent representation of (\ref{JJOO-gen}) involving the conjugated
tensor $\widebar{\mathcal{M}}_{\{\dot{\alpha}\}\{\dot{\beta}\}\{\dot{\gamma}\}\{\dot{\delta}\}}$ carrying anti-chiral indices. 

We note in passing that the correlation function involving the supersymmetry current \p{diffS} and its conjugate can be derived by the same method and the expression has a form similar to \p{JJOO-gen}
\begin{align}\label{}
 \vev{ \Psi_{\a\b\da}^{a'}(1)\, \bar\Psi^a_{\db\dot\gamma\gamma}(2)\, O(3)\, O(4)}^{\nr} =  (\partial_{x_1})_{\da}^{\delta_1}(\partial_{x_2})_{\db}^{\delta_2} (\partial_{x_2})_{\dot\gamma}^{\delta_3}\ \left[\cM^{aa'}_{\a\b\gamma\delta_1\delta_2\delta_3}  \Phi(u,v)\right]
\end{align}
with $\ep^{\a\delta_1} \cM^{aa'}_{\a\b\gamma\delta_1\delta_2\delta_3}=\ep^{\gamma\delta_3} \cM^{aa'}_{\a\b\gamma\delta_1\delta_2\delta_3}=0$.

This concludes our general discussion of the method for reconstructing the complete correlation function of four energy-momentum supermultiplets, starting from its 
lowest bosonic component. We have given a general and very compact formula for the super-correlation function, Eq.~\p{G4-main}. We have shown several examples 
of how to extract various component correlators. The examples are limited to the case where $\q_i=\bq_i=0$ at two of the four points. Including the dependence on all 
four sets of odd variables, in order to obtain components like $\vev {T(1)T(2)T(3)T(4)}$ with four energy-momentum tensors, considerably complicates the algebra. We 
should mention that in the special case of $\cN=4$ SYM the component at $(\q)^4$ in the expansion \p{T} is the (chiral on-shell) Lagrangian ${\cal L}$ of the theory, and 
similarly for the anti-chiral $\tilde{\cal L}$ at $(\bq)^4$.  In this case one can obtain the component $\vev {{\cal L}(1){\cal L}(2)\tilde{\cal L}(3)\tilde{\cal L}(4)}$ which is 
known to be the AdS/CFT dual of the amplitude of dilatons and axions in AdS${}_5\times S^5$ supergravity \cite{D'Hoker:1999pj}. In the past this component had been 
computed  in \cite{Drummond:2006by} using a different method based on $\cN=2$ supersymmetry. The result of Ref.~\cite{Drummond:2006by} is surprisingly simple, 
so it would be interesting to obtain it by our new method.

\section{Charge flow correlations}\label{s4}

In this section, we explain how the particular class of correlation functions~(\ref{hatG4}), discussed in the previous section, is connected to recent work 
\cite{Belitsky:2013xxa,Belitsky:2013bja,Belitsky:2013ofa} on the so-called charge-flow correlations.

The charge-flow correlations measure the flow
of various quantum numbers (e.g., $R-$charge, energy) in the final states created from the vacuum by a particular source. Schematically, these correlations are written in the form \begin{align}\label{F-def}
\vev{\mathcal D_{s}(n) \mathcal D_{s'}(n')}_q = {(q^2)^{s'-1}(qn')^{s-s'} \over  (nn')^{s+1} }  \cF_{ss'} (z)\,,
\end{align}
and depend only on the total four-momentum $q^\mu$ transferred by the source and the dimensionless scaling variable $z=q^2 (nn') /(2(qn)(qn'))$ which is constructed from the kinematical data of the process. More precisely, the {\it flow operators} $\mathcal{D}_s(n)$ can be interpreted as `detectors' located at spatial infinity that measure the flow of a particular quantum number (labelled by the integer Lorentz spin $s$) per unit angle in the direction of the light-like four-vector $n_\mu$. The variable $z$ corresponds to the angle between the two detectors $\mathcal{D}_{s}(n)$ and $\mathcal{D}_{s'}(n')$. The event shape function $\cF_{ss'} (z)$ can be understood as a differential distribution of the charges measured by the two detectors.

\subsection{Basics of charge flow correlations}

The double flow correlations can be defined in terms of correlation functions   
\begin{align}\label{G-flow}\notag
& \vev{\mathcal D_{s}(n) \mathcal D_{s'}(n')}_q =  i^{s+s'}  \sigma_{\text{tot}}^{-1} \int d^4 x_{34} \e^{iq x_{34}} G_{ss'}  \,,
\\[2mm]
& \  G_{ss'}=\vev{  \mathcal D_{s}(n)\, \mathcal D_{s'}(n')\, O(x_3,y_3)\, O(x_4,y_4)}_{W}\,,
\end{align}
where the additional factor $i^{s+s'}$ is inserted
to ensure reality for $\vev{\mathcal D_{s}(n) \mathcal D_{s'}(n')}_q$ (see Eq.~\re{calF} below) and the Fourier transform
is performed with respect to $x_{34}=x_3-x_4$.
 
The charge flow operators $\mathcal D_{s}(n)$ and $\mathcal D_{s'}(n')$ depend on light-like vectors, $n$ and $n'$ (with $n^2=n'{}^2=0$), 
and are defined in terms of local operators (including conserved currents) of spin $s$ and $s'$, respectively. Particularly important 
examples of $\mathcal D_{s}(n)$  corresponding to $s=0,1,2$  are 
\begin{align}\notag\label{flow}
& \mathcal  O(n,y)  =  (n\bar n) \int_{-\infty}^\infty  d\tau \,  \lim_{r\to\infty} r^2 \, \OO (r n  + \tau \bar n,y) \,, 
\\\notag
& \mathcal Q^{aa'}(n,y) =     \int_{-\infty}^\infty  d\tau  \lim_{r\to\infty} r^2 \,( J_-  )^{aa'}(r n  + \tau \bar n,y)  \,,
\\
& \mathcal E(n) =  {1 \over  (n\bar n)} \int_{-\infty}^\infty  d\tau  \lim_{r\to\infty} r^2 \,T_{--}(r n  + \tau \bar n)\,,
\end{align}
where $\bar n$ is an auxiliary light-like vector, $\bar n^2=0$ and $(n\bar n)\neq 0$, \footnote{The expressions on the right-hand sides of \re{flow} involve the 
two light-like vectors $n$ and $\bar n$, but the dependence on the latter is redundant. We can exploit this fact 
to put  $\bar n^\mu=(1,0,0,-1)$.} while
\begin{align}\label{4.3}
&J_-(x)\equiv {\bar n^\mu J_\mu(x)} \,,\qquad\qquad T_{--}\equiv  {\bar n^\mu\bar n^\nu T_{\mu\nu}(x)}  
\end{align} 
are the light-cone components of the $R-$current and the energy-momentum tensor, respectively.  The total cross section $\sigma_{\text{tot}}$ in \re{G-flow} is  defined as
\begin{align}\label{4.4}
\sigma_{\text{tot}}(q)=\int d^4x_{3}\,e^{iqx_{3}}\vev{O(x_3,y_3)\,O(0,y_4)}_W\,.
\end{align}
The subscript $W$ in the second line of (\ref{G-flow}) and in \p{4.4} indicates the Wightman four- and two-point functions computed in four-dimensional 
Minkowski space of signature $(+,-,-,-)$. They can be obtained from their Euclidean counterparts  through an analytic continuation.  

Our discussion will concern only the anomalous 
part of the correlation function \p{G-flow} because the rational part gives rise to contact terms. To simplify the notation, from now on we drop the superscript `$\nr$'. 

\subsection{Change of coordinates}

The expression for the flow operators \re{flow} can be simplified by a change of coordinates $x^\mu\mapsto z^\mu$, as proposed in~\cite{Cornalba:2007fs,Hofman:2008ar}. Indeed, it is well known that four-dimensional Minkowski space can be embedded as a light-like surface in six-dimensional projective space $\eta^M$ (with $M=0,1,2,3,5,6$)
 \begin{align}\label{1}
\eta^2 = (\eta^0)^2 - (\eta^1)^2  - (\eta^2)^2  - (\eta^3)^2  - (\eta^5)^2 + (\eta^6)^2 =0\,.
\end{align}
The two sets of Minkowski coordinates, $x^\mu$ and $z^\mu$ (with $\mu=0,1,2,3$), are defined as  
\begin{align}\label{2}\notag
x^\mu &=\left( \frac{\eta^0}{\eta^5 + \eta^6}\,, \frac{\vec\eta}{\eta^5 + \eta^6} \,,\frac{\eta^3}{\eta^5 + \eta^6}\right),
\\
z^\mu &=\left( -\frac{\eta^6}{\eta^0 + \eta^3}\,,  \frac{\vec\eta}{\eta^0 + \eta^3}\,,  -\frac{\eta^5}{\eta^0 + \eta^3} \right),
\end{align} 
where $\vec{\eta}=(\eta_1,\eta_2)$. They are related by a rotation in the six-dimensional embedding space 
or equivalently by a conformal transformation in Minkowski space. 

The relation between the coordinates \re{2} is
particularly simple in the  light-cone coordinates 
defined as
\begin{align}
x_{\a\da} = x^\mu (\sigma_\mu)_{\a\da}= \left[\begin{array}{ll}x^+ & \bar x \\ x & x^- \end{array}\right]\,,\qquad x^\pm =x^0\pm x^3\,,\qquad x = x^1+i x^2 \,, 
\end{align}
so that $(x^\mu)^2 = \det\| x^{\a\da}\| =  x^+x^- - \vec x{\,}^2$ with $\vec x=(x^1,x^2)$.
The new coordinates are given by
\begin{align}\label{5}
&z^+=-\frac1{x^+}\,,\qquad \qquad z^- = x^--\frac{\vec x{\,}^2}{x^+}\,, \qquad \qquad  \vec z= \frac{\vec x}{x^+}\, .
\end{align}
It is easy to verify that this change of variables  amounts to a conformal transformation of the coordinates $x^\mu$ with a simple  
weight factor, $d z^\mu d z_\mu=d x^\mu dx_\mu/(x^+)^2$. 

In the special case of the flow operators \re{flow}, with $x^\mu = r n^\mu + \tau\bar n^\mu$, the new coordinates for 
$r\to\infty$ are given by 
\begin{align}
&z^+ =  0\,,\qquad \qquad z^- = x^- - r {\vec n^2\over n^+}\,,\qquad \qquad \vec z = {\vec n \over n^+}\,,
\end{align}
where  we took into account that $\bar n^\mu=(1,0,0,-1)$ has  only one non-zero light-cone coordinate, $\bar{n}^+=0$ and $\bar n^-=2$.
Since the operators $\OO $, $J_\mu$ and $T_{\mu\nu}$ in \re{flow} transform covariantly under conformal transformations, the flow operators 
in the new coordinates have the following form
\begin{align}\label{15'}
 &{\cal O}(\vec z, y) ={1\over 2n^+} \int^\infty_{-\infty} dz^- O(0^+,z^-, \vec z\ ; y)\,,\nt
 &{\cal Q}^{aa'}(\vec z, y) = {1\over (n^+)^2}\int^\infty_{-\infty} dz^- J^{aa'}_-(0^+,z^-, \vec z\ ;y)\,,\nt
 &{\cal E}(\vec z) =  {1\over (n^+)^3}\int^\infty_{-\infty} dz^- T_{--}(0^+,z^-, \vec z) \,,
\end{align}
where $0^+$ stands for $z^+=0$ and $J_-$ and $T_{--}$ were defined  in \p{4.3}.  
Compared with \re{flow}, the dependence of the flow operators  on $n^\mu$  in the new coordinates enters through the 
two-dimensional vector $\vec z=\vec n/ n^+$. Also, most importantly, the limit $r\to\infty$ in \re{flow} is now replaced by setting $z^+=0$, which is technically easier to implement. 

Denoting the flow operators \re{15'} as $\mathcal D_{s}(\vec z)$ (for $s=0,1,2$), we find that
the correlation function \re{G-flow} admits the following representation in the new coordinates
\begin{align}\label{G-z}
G_{ss'}=(z_3^+ z_4^+)^2\vev{ \mathcal D_{s}(\vec z) \mathcal D_{s'}(\vec z{\,}')\OO (z_3,y_3)  \OO (z_4,y_4)} \,,
\end{align}
where $(z_3^+ z_4^+)^2$ arises as the conformal weight of the operators $\OO (z_3,y_3)\OO (z_4,y_4)$ under the change of
variables. For general values of the spins $s$ and $s'$, \re{G-z} is given by the four-point
correlation function \re{JJOO-gen} integrated over the light-cone coordinates of the two currents
\begin{align}\label{minus-int}\notag
 G_{ss'} {}& ={(z_3^+ z_4^+)^2\over (n^+)^{s+1}(n'{}^+)^{s'+1}} \int^\infty_{-\infty} dz_1^- dz_2^- 
 \\[2mm]
{}&\qquad \times  \vev{J_{\underbrace{\scriptstyle -\ldots -}_{s}}(0^+,z_1^-,\vec z_1)  J_{\underbrace{\scriptstyle -\ldots -}_{s'}}(0^+,z_2^-,\vec z_2)  \OO (z_3,y_3)   \OO (z_4,y_4)} \,,
\end{align}
where $\vec z_1=\vec n/n^+$ and $\vec z_2=\vec n'/n'{}^+$.
 Notice that $G_{ss'}$ only involves the minus components
of the currents defined by
\begin{align}\label{minus}
J_{-\ldots -} = \bar n^{\mu_1}\dots \bar n^{\mu_s} J_{\mu_1\dots \mu_s} = \frac1{2^s}\bar n^{\da_1\a_1} \dots \bar n^{\da_s\a_s}
J_{\a_1\da_1\dots \a_s\da_s}\,.
\end{align} 
As we will see in a moment, this property makes the new coordinates particularly useful for
analyzing the properties of $G_{ss'}$.

\subsection{Light-cone superfield}
\label{SectionLCsuperfield}

We recall that the conserved currents $J_\mu(z)$ and $T_{\mu\nu}(z)$ appear as components in the expansion of the 
energy-momentum  supermultiplet \re{T-gen}. As we explained in Sect.~\ref{desc},  to extract their correlation functions from the correlation function of the superfields $\cT(z,\q,\bq,y)$, we have to subtract the contribution of the conformal descendants involving total derivatives,  like $\pa_z \pa_y O(z,y)$ for the $R-$symmetry current $J_\mu$, Eq.~\re{J}. 

To compute the correlation function in \re{minus-int} we do not need the whole expression for the superfield  \re{T-gen}. 
It is sufficient to retain only the terms containing the minus components of the currents. Since in \re{T} the Lorentz indices of the currents
are contracted with $\theta \sigma_\mu  \bar \theta$, this can be done by 
imposing the additional condition on the Grassmann variables 
\begin{align}\label{zero}
 \theta^{a \alpha} (\sigma_\mu)_{\a\da}  \bar \theta_{a'}^{\dot\alpha}  = 0\,, \qquad \text{for $\mu\neq -$}\,.
 \end{align}
Moreover,  in this case the contribution  of  the conformal descendants   is  proportional to total derivatives with respect to the light-cone
coordinate $z^-$, like $\pa_{z^-} \pa_y O(z,y)$, and it vanishes after the integration on the right-hand side of \re{minus-int}. This allows us to 
safely neglect the  descendants in \re{minus-int}.
   
To solve \re{zero}, we use the projectors 
$\frac14 \sigma^{\pm} \sigma^\mp$ to decompose $\theta^{a \alpha}$ and $\bq_{a'}^{\da}$ into  sums of two components,  
\begin{align}\label{theta-minus} \notag
& \theta^{a \a} 
= \ft14(\sigma^{+} \sigma^- + \sigma^{-} \sigma^+) \theta^{a} 
\equiv  (\q^{a,-},-\q^{a,+})\,,
\\
&\bar\theta_{a'}^{\da}
= \ft14( \sigma^{+}\sigma^- + \sigma^{-}\sigma^+) \bar\theta_{a'} \equiv \left(\begin{array}{l} \bq_{a'}^{-} \\[2mm] \bq_{a'}^{+} \end{array}\right).
\end{align}
Substitution of  \re{theta-minus} into \re{zero}  yields the relation $\theta^{a,+}  \bar\theta_{a'}^{\,+} =0$ which has obviously two solutions. 
The expressions for the superfield $\cT$ evaluated on the shell of these solutions differ by the contribution of half-integer operators.
Since here we are only considering flow operators of integer spin, we can replace the above constraint by a stronger one, i.e., 
\begin{align}\label{4.15}
 \theta^{a,+} =\bar\theta_{a'}^{\,+} = 0\,.
\end{align}
The resulting light-cone superfield $\cT_-(i) \equiv \cT(z_i,\q_i^-,\bq_i^-,y_i)$ takes the following form
\begin{align}\label{T-minus}
\cT_-(i) =  \OO (z_i)+ \dots + (\q_i^-)^a  (\bq_i^-)_{a'} (J_{-} (z_i))^{a'}_a + \dots +  (\q_i^-)^2 (\bq_i^-)^{2} T_{--}(z_i)\,,
\end{align}
where $ (\q_i^-)^2= \prod_{a} (\q_i^-)^a$ and similarly for $(\bq_i^-)^{2}$.
Substituting $\theta_i^4=(\theta_i^+)^2 (\theta_i^-)^2$ in \p{G4-main} and replacing  $x_i^\mu$ by the new 
coordinates $z_i^\mu$, we obtain 
\begin{align}\label{G4-main1}\notag
& \vev{\cT_-(1) \cT_-(2)  \OO (3) \OO (4)}  =  \lr{\frac{y^2_{13}}{\hat z^2_{13}}\frac{y^2_{24}}{\hat z^2_{24}}}^2
\\[2mm]
& \qqquad\qquad \times    \bar S^4\bar S'{}^4Q^4Q'{}^4
\left[    {(\q_1^+)^2(\q_1^-)^2  (\q_2^+)^2(\q_2^-)^2 \q_3^4 \q_4^4 \over (z_{13}^2 z_{24}^2 y_{13}^2 y_{24}^2)^2}{\Phi(u,v) \over uv}  \right]\bigg|_{\theta^+_{1,2}=\bar\theta_{1,2}^+=\theta_{3,4}=\bar\theta_{3,4}=0}\,,
\end{align}
with $\hat z_{ij}$ defined as in \p{susy-pro}. 
 By construction, the expansion of \re{G4-main1} in powers of $\theta^-_{1,2}$ and 
$\bar\theta_{1,2}^-$ generates the correlation functions involving the minus components of the currents at points $1$ and $2$
and the half-BPS operators $\OO $ at points $3$ and $4$. As follows from \re{JJOO-gen}, such correlation functions have 
a very special form. 

Let us consider the simplest example of the correlation function \re{G-simple}, involving a single insertion of the current. According to \p{minus}, we have to contract \re{G-simple} with $\bar n^{\da_1\a_1} \dots \bar n^{\da_s\a_s}/2^s$. Taking into account the identities (see Appendix \ref{App:A})
\begin{align}
&\frac12 \bar n^{\da\a} = \left[\begin{array}{cc} 1 & 0 \\ 0 & 0\end{array}\right]\,,&&
\frac12 (\partial_{z})^\b_{\da}\,\bar n^{\da\a} 
= \left[\begin{array}{ll} -\partial_{\bar z} & 0 \\ - \partial_{z^-} & 0\end{array}\right]\,, 
\end{align}
with $\bar z = z^1 - i z^2$, we find from \re{G-simple}
\begin{align}
 \vev{J_{\underbrace{\scriptstyle -\ldots -}_{s}}(1) \OO(2) \OO(3) \OO(4)}  =
(\partial_{\bar z_1})^s \big[ \cM_{s}\Phi(u,v)] + \dots, 
\end{align} 
where $\cM_{s} \equiv \cM_{\a_1\b_1\dots \a_s\b_s}$ with $\a_i=\b_i=1$
and the dots denote terms involving total derivatives with respect to $z_1^-$. As was explained in the beginning of this subsection,
such terms produce vanishing contributions in the integrated correlation function \re{minus-int} and can be safely neglected.
Repeating the same analysis for the correlation function \re{JJOO-gen}, we obtain
\begin{align}\label{short}
\vev{J_{\underbrace{\scriptstyle -\ldots -}_{s}}(1) J_{\underbrace{\scriptstyle -\ldots -}_{s'}}(2) \OO(3) \OO(4)}  =
(\partial_{\bar z_1})^s(\partial_{\bar z_2})^{s'} \big[ \cM_{ss'}\Phi(u,v)] + \dots, 
\end{align}\\[-3mm]
where the dots denote terms with total derivatives with respect to $z_1^-$ and $z_2^-$. Here $\cM_{ss'}$ is a (complicated)
rational function depending on the coordinates of all 4 points. Its explicit form is fixed unambiguously by \re{G4-main1} but, as
we have shown in the previous section, the calculation can be very involved. 

The power of the equation \re{short} lies in the fact that only one term in the expansion, namely the one which contains the maximal number of 
anti-holomorphic derivatives $(\partial_{\bar z_1})^s(\partial_{\bar z_2})^{s'} \Phi(u,v)$, is sufficient to determine $\cM_{ss'}$.
In the expansion of the super correlation function \re{G4-main1}, this term is accompanied by a product of Grassmann 
variables and reads, schematically,
\begin{align}\label{short1}
 \vev{\cT_-(1) \cT_-(2)  \OO (3) \OO (4)}  =\sum_{s,s'\ge 0}
 (\theta_1^- \bar\theta_1^-)^s (\theta_2^- \bar\theta_2^-)^{s'} \cM_{ss'}  (\partial_{\bar z_1})^s(\partial_{\bar z_2})^{s'} 
    \Phi(u,v) +\dots\,,
\end{align}\\[-3mm]
where the $SU(2)$ indices of $\theta_i^{a,-}$ and $\bar\theta_{i,a'}^-$ are contracted with those of $ \cM_{ss'}$. Here the
dots denote terms with derivatives distributed between $\Phi(u,v)$ and $\cM_{ss'}$.

\subsection{Gauge fixing}

As was shown in Sect.~\ref{sect:gauge}, the calculation of \re{G4-main1} can be significantly simplified by an appropriate choice of 
additional conditions on the bosonic coordinates. 

The gauge \re{gauge} cannot be employed in \re{G4-main1} 
due to singularities in the change of variables \re{5}. However, we can impose the following weaker condition
 on the coordinates of the half-BPS operators at points $3$ and $4$
\begin{align}\label{gauge1}
 y_{3,aa'} = 0\,,\qquad   y_{4,aa'} \to \infty\,,\qquad \vec z_3=\vec z_4=z_4^-=0\,,\qquad  z_4^+\to\infty\,,
\end{align}
 where $\vec z_i=(z_i^1,z_i^2)$. Moreover, since \re{minus-int} involves the currents at zero values of the
`$+$'--light-cone components, we impose analogous condition on the coordinates of the superfields 
 at points $1$ and $2$,
 \begin{align}\label{gauge3}
z_1^+=z_2^+=0\,.
\end{align}
The main advantage of this gauge is that it allows us to simplify \p{G4-main1} by eliminating 
the dependence of $\theta_{3,4}$ and $\bar\theta_{3,4}$ in \re{G4-main1}. 

We find
the following result for the correlation function \re{G4-main1} in the gauge \re{gauge1} and \re{gauge3}  (for details see Appendix \ref{App:C})
\begin{align}\label{G4-main3} 
& \vev{\cT_-(1) \cT_-(2)  \OO (3) \OO (4)}  =   
   {  (y^2_{4})^2 \over (z_{12} \bar z_{12}  z_3^-)^4(z_3^+z_4^+)^2}  \,
{\rm S}_{1}^2\, {\rm S}_{2}^2
\left[  (\q_1^-)^2 (\q_2^-)^2 F(u,v)  \right] +\dots\,,
\end{align}
where the dots have the same meaning as in \p{short1} and
\begin{align}\label{C4}
F(u,v) =  {u^3\over v}\Phi(u,v)\,.
\end{align}
Here  ${\rm S}_{i}^2=\prod_{a'=1,2} {\rm S}_{i,a'}$ stands for the product of the linear differential operators acting on the coordinates
at points $1$ and $2$,
\begin{align}\notag\label{rmS1}
& {\rm S}_{1,a'} 
=-\bar z_{12} z_3^+ (\bq_{1}^-)_{a'} {\partial_{\bar z_1}} + \sum_{i=1,2} (\Omega_{1i})_{a'}{}^a {\partial_{\q_i^{a,-}}}\,,
\\
& {\rm S}_{2,a'} 
=  \bar z_{12} z_3^+ (\bq_{2}^-)_{a'} {\partial_{\bar z_2}} + \sum_{i=1,2} (\Omega_{2i})_{a'}{}^a {\partial_{\q_i^{a,-}}}
\end{align}
with matrices $\Omega_{1i}$ and $\Omega_{2i}$ given by
\begin{align}\notag\label{Omega}
& \Omega_{11} = y_1 (z_1 \bar z_2- z_3^+ z_3^-)\,, && \Omega_{12} = y_1 (z_2\bar z_2-z_3^+z_3^-)+y_{12} z_2^-z_3^+\,,
\\[2mm]
& \Omega_{21} = y_2 (z_1\bar z_1 -z^+_3z^-_3)-y_{12} z^-_1 z^+_3 \,,&& \Omega_{22} = y_2(z_2 \bar z_1-z^+_3z^-_3)\,.
\end{align}
Notice that if we set $(y_{12})_{a'}^a=0$, the generators \re{rmS1} cease to depend on $z_1^-$ and $z_2^-$. As we show
in the next subsection, this restriction of the $y-$dependence corresponds to choosing a particular irreducible representation of  $SU(4)$. 
It plays a crucial role in establishing relations between various charge-flow correlations.  

We would like to stress that  Eq.~(\ref{G4-main3}) refers only to the contributions  to the correlation function  
of the form  \re{short1},
for which all the space-time derivatives act  on the function $\Phi(u,v)$, or equivalently $F(u,v)$. The  remaining terms (including those
involving total derivatives with respect to $z_1^-$ and $z_2^-$) are represented by the dots in \re{G4-main3}.
Arriving at \re{G4-main3}, we replaced $\hat z_{13}^2\to z_{13}^2$ and $\hat z_{24}^2\to z_{24}^2$ since the remaining terms 
produce contributions which involve the product of Grassmann variables $(\theta_i^- \bar\theta_i^-)$  but which  is not 
accompanied by derivatives $\partial_{\bar z_i}$. Also, we took into account that $z_{i4}^2 = -z_4^+ z_i^-$ and $z_{12}^2=-z_{12} \bar z_{12}$ 
in the gauge \re{gauge1} and \re{gauge3}.
 
Replacing the generators $\rm S_1$ and $\rm S_2$ in \re{G4-main3} by their explicit expressions \re{rmS1}, we can expand
the correlation function \re{G4-main3} in powers of the Grassmann variables and match the result with \re{short1} to identify
$\cM_{ss'}$. We can then use these functions to compute \re{short}
in the gauge \re{gauge1} and \re{gauge3}. In the next subsection, we perform this calculation for a particular $SU(4)$ component
of the correlation function \p{G4-main3}.
  
\subsection{Top Casimir components}

The two currents in \re{short} belong to some irreducible representations of the $R-$symmetry group  $SU(4)$, e.g., $\mathbf{15}$ for 
the $R-$current and $\mathbf{1}$ for the energy-momentum tensor. These representations, denoted 
by $\mathbf{A}_1$ and $\mathbf{A}_2$,  are encoded in the dependence of the matrix $\mathcal M_{ss'}$ in \p{short} on $(y_i)_{a'}^a$. More precisely,
we can decompose $\mathcal M_{ss'}$ into its irreducible components
\begin{align}\label{R-dec}
\mathcal M_{ss'}(y,z) = \sum_R \cY_{ss';R}(y) \mathcal M_{ss';R}(z)\,,
\end{align}
where the (finite) sum runs over the overlap of $SU(4)$ representations $R$ in the tensor products $\mathbf{A}_1\times\mathbf{A}_2$ and
$\mathbf{20' \times 20'  =  1+15+20'+84+105+175}$. Here $\cY_{ss';R}(y)$ are polynomials in $y_i$ defined as eigenfunctions
of the quadratic Casimir of  $SU(4)$ for the representations $R$ and the coefficients $\mathcal M_{ss';R}$ are  independent
of $y_i$. Explicit expressions for $\cY_{ss';R}(y)$ can be found in \cite{Belitsky:2013bja}.

In what follows we shall concentrate on the particular component in \re{R-dec} with the maximal value of the $SU(4)$ quadratic
Casimir. We shall refer to it as the top Casimir component. The reason for this is that, as we show below, for different choices of the currents in  \re{short}  the corresponding coefficients 
$\mathcal M_{ss';R}$ have a remarkably simple universal form. The simplest way to select this special component is to set $y_1=y_2$.
In this case, as explained in Appendix~\ref{slyy}, all but one $\cY_{ss';R}(y)$ vanish and the correlation function \re{short} receives contributions from the top Casimir
component only\footnote{The only exception is the case with two $R-$currents, i.e. $s=s'=1$ and  $\mathbf{A}_1=\mathbf{A}_2=\mathbf{15}$, where an additional symmetrization of the 
$SU(2)$ indices is required, see Eq.~\re{JJ} below.}.
  
Let us now examine  \re{G4-main3} for $y_1=y_2$. We observe that the expression for the generators \re{rmS1} simplifies due to
\begin{align}\label{omega}
(\Omega_{ij})_{a'}{}^a = (y_1)_{a'}{}^a\omega_{ij} \,,\qquad \ \omega=\left[\begin{array}{cc}z_1 \bar z_2- z_3^+ z_3^- & z_2\bar z_2-z_3^+z_3^- 
\\[2mm] z_1\bar z_1 -z^+_3z^-_3 & z_2 \bar z_1-z^+_3z^-_3 \end{array}\right],
\end{align}
with $\det \omega=z_{12} \bar z_{12} z_3^+ z_3^-$.
Then, using the integral representation ${\rm S}_{i}^2 = \int d^2 \epsilon \e^{\epsilon^{a'} {\rm S}_{i a'}}$ we find after some algebra
\begin{align}\notag
& {\rm S}_{1}^2\, {\rm S}_{2}^2
\left[  (\q_1^-)^2 (\q_2^-)^2 F(u,v)  \right] = (y_1^2)^2 (\det \omega )^2
\\[2mm]
 & \qqquad \times{}  
   \exp\left\{  {\bar z_{12} z_3^+\over \det \omega}\left[
\lr{\omega_{22}  \q_1^-- \omega_{21} \q_2^- }y_1^{-1}\bq_1^- \partial_{\bar z_1} - 
\lr{\omega_{11} \q_2^- -\omega_{12} \q_1^-}y_1^{-1}\bq_2^- \partial_{\bar z_2} \right] \right\} F(u,v)\,.
\end{align}
It is easy to see that the expansion of this expression in powers of the Grassmann variables has the expected form \re{short1},
\begin{align}\notag\label{top}
  {\rm S}_{1}^2\, {\rm S}_{2}^2
\left[  (\q_1^-)^2 (\q_2^-)^2 \Phi(u,v)  \right]  & =(y_1^2)^2 (z_{12}\bar z_{12} z_3^+ z_3^-)^2 F(u,v) + \dots 
\\[2mm]
& 
 +(\bar z_{12} z_3^+)^4 (\q_1^-)^2 (\q_2^-)^2(\bq_1^-)^2 (\bq_2^-)^2 \lr{\partial_{\bar z_1} \partial_{\bar z_2}}^2 F(u,v)\,.
\end{align}
Substituting this relation into  \re{G4-main3} and matching the various terms of the expansion with \re{short1}, we can identify the coefficient  $\mathcal M_{ss';R}$ corresponding to the top Casimir component in \re{R-dec}.

Let us start with the lowest component of \re{top}. Its contribution to \re{G4-main3} looks as
\begin{align} \label{low-exp}
& \vev{\cT_-(1) \cT_-(2)  \OO (3) \OO (4)}_{\q=\bq=0} = 
   {(y_1^2y_4^2)^2 \over (z_{12} \bar z_{12} z_3^- z_4^+)^2} F(u,v) \,.
\end{align}
This expression should be matched with the correlation function
$\vev{\OO (1)\OO (2) \OO (3) \OO (4)}$ involving the lowest component of the superfields $\cT_-(1)$ 
and $\cT_-(2)$, Eq.~\re{T-minus}. Indeed, replacing $x_i$ with $z_i$ in \re{G4-loop} and imposing the gauge condition \re{gauge1},  
we find for $y_1=y_2$, i.e. for the top Casimir $SU(4)$ channel $R=\mathbf{105}$, 
\begin{align} \label{G4-z}
\vev{\OO (1)\OO (2) \OO (3) \OO (4)}_{\mathbf{105}}
=  &  \frac{(y_{1}^2y_{4}^2)^2  }{(z_{12}^2 z_{34}^2)^2} F(u,v)\,.
\end{align}  
Taking into account that $z_{12}^2 z_{34}^2=z_{12} \bar z_{12} z_3^-z_4^+$ in the gauge  \re{gauge1} and  \re{gauge3}, we
find agreement with \re{low-exp}.  
  
Similarly, we identify the contribution to \re{G4-main3} coming from the highest component in \re{top},
\begin{align} \label{top-exp} 
& \vev{\cT_-(1) \cT_-(2)  \OO (3) \OO (4)}   =  
   { (y_4^2)^2 ( z_3^+)^2 \over (z_{12}   z_3^-)^4 (z_4^+)^2}  (\q_1^-)^2 (\q_2^-)^2(\bq_1^-)^2 (\bq_2^-)^2 \lr{\partial_{\bar z_1} \partial_{\bar z_2}}^2 F(u,v) + \ldots  \,.
\end{align}
It should be confronted with the correlation function involving the  $T_{--}$ components of the energy-momentum tensors (see Eq.~\re{T-minus}). Comparing
\re{top-exp} with \re{short1}, we identify the corresponding function $\cM_{s=2,s'=2}$ and apply \re{short} to obtain
\begin{align}\label{TT-1}
\vev{T_{--}(1)T_{--}(2)\OO (3) \OO (4)}_{\mathbf{1}} =   
   {(y_4^2)^2 ( z_3^+)^2 \over (z_{12}   z_3^-)^4(z_4^+)^2}  \lr{\partial_{\bar z_1} \partial_{\bar z_2}}^2    
  F(u,v) +\dots\,.
\end{align}
Here the dots denote terms involving total derivatives with respect to the minus space-time components. We recall that we have systematically neglected such terms from the start. The subscript $\mathbf{1}$ indicates the unique $SU(4)$ channel in this case, the singlet representation of $SU(4)$. 
  
It is straightforward to extend the analysis to the remaining components of the expansion  \re{top}. Below we present the results
for various correlation functions involving the $R-$current and energy-momentum tensor.
We start with a single current insertion
\begin{align}\notag\label{JO-TO}
& \vev{J_{-}^{a'a} (1) \OO (2)\OO (3) \OO (4)}_{\mathbf{175}} =  { y_1^2 (\ty_1)^{a' a} (y_4^2)^2 \over (z_{12} z_3^-)^3 (z_4^+)^2} \partial_{\bar z_1} \bigg[{(z_2 \bar z_1 - z_3^+ z_3^-) \over \bar z_{12}^2}F(u,v)\bigg],
\\[2mm]
& \vev{T_{--}(1)\OO (2)\OO (3) \OO (4)}_{\mathbf{20'}} =  {y_2^2(y_4^2)^2  \over (z_{12} z_3^-)^4 (z_4^+)^2} \partial^2_{\bar z_1} \bigg[{(z_2 \bar z_1 - z_3^+ z_3^-)^2 \over \bar z_{12}^2} F(u,v)\bigg].
\end{align}  
We verified that these relations are in agreement with \re{onecu} and \re{T1-con}, respectively, for $y_1=y_2$ after we impose the gauge conditions \re{gauge1} and \re{gauge3}. The second relation in \p{JO-TO} does not depend on $y_1$ and holds for arbitrary $y_2$. For the double $R-$current insertion we find
\begin{align}\notag\label{JJ}
& \vev{J_{-}^{a'a} (1)J_{-}^{b'b} (2)\OO (3) \OO (4)}_{\mathbf{20'+ 84}} 
 \\[2mm]
& \qquad=  {(y_4^2)^2 \over (z_{12} z_3^-)^4 (z_4^+)^2} \partial_{\bar z_1} \partial_{\bar z_2}\bigg[\lr{\omega_{11} \omega_{22} (y_1)^{a'a}(y_1)^{b'b} - \omega_{12} \omega_{21} (y_1)^{a'b}(y_1)^{b'a}  }{F(u,v)\over \bar z_{12}^2} \bigg],
\end{align}
with $\omega$  defined in \re{omega}.
Here the condition $y_1=y_2$ leaves two (and not one as in the other cases) irreducible representations in 
\re{R-dec}, the $\mathbf{20'}$ and the $\mathbf{84}$. The latter corresponds to the top Casimir contribution. It can be singled out by decomposing
$(y_1)^{a'a}(y_1)^{b'b}$ and $(y_1)^{a'b}(y_1)^{b'a}$ into irreducible components with respect to the   
little group $SU(2)\times SU(2)'$ of the harmonic coset (see Appendix \ref{slyy}), e.g.,
\begin{align}\notag
& (y_1)^{a'a}(y_1)^{b'b} = (y_1)^{(a'a}(y_1)^{b')b} - \ft12 \epsilon^{a'b'}\epsilon^{ab} y_1^2\,.
\end{align}
Then, we find from \re{JJ} 
\begin{align} \label{JJ-irr}
& \vev{J_{-}^{a'a} (1)J_{-}^{b'b} (2)\OO (3) \OO (4)}_{\mathbf{84}} 
 =  {(y_1)^{(a'a}(y_1)^{b')b} (y_4^2)^2  z_3^+ \over (z_{12} z_3^-)^3 (z_4^+)^2} \partial_{\bar z_1} \partial _{\bar z_2}\bigg[    {F(u,v)\over \bar z_{12} } \bigg].
\end{align}
Finally, for the insertion of an $R-$current and an energy-momentum tensor we have 
\begin{align}\label{TJ-top}
 \vev{T_{--}(1)J_{-}^{a'a} (2)\OO (3) \OO (4)}_{\mathbf{15}} =  {(y_2)^{a' a}(y_4^2)^2 z_3^+ \over (z_{12} z_3^-)^4  (z_4^+)^2} \partial^2_{\bar z_1} \partial _{\bar z_2}\bigg[{(z_2 \bar z_1 - z_3^+ z_3^-)  \over \bar z_{12} } F(u,v)\bigg].
\end{align}
We would like to emphasize that the expressions for the correlation functions derived in this subsection are valid up to terms
involving total derivatives with respect to $z_1^-$ and $z_2^-$. Such terms do not contribute to the charge-flow correlations
\re{minus-int}.
     
\subsection{Integrated correlation functions}
  
Let us now apply the results obtained in the previous subsection to compute  \re{minus-int}. 

Comparing relations \re{top-exp} -- \re{TJ-top} with the general expression \re{short} we observe the following remarkable
feature of the top Casimir component of each correlation function: the corresponding functions $\cM_{ss'}$ do not depend
on the light-cone coordinates $z_1^-$ and $z_2^-$.~\footnote{As was already mentioned, this property follows from the fact that the 
$\Omega-$matrix \re{Omega} ceases to depend on  $z_1^-$ and $z_2^-$ for $y_1=y_2$.}
As a consequence, substituting \re{short} into \re{minus-int} we find for the top Casimir component
\begin{align}\label{minus-int1}
 G_{ss',{\rm top}}={(z_3^+ z_4^+)^2\over (n^+)^{s+1}(n'{}^+)^{s'+1}}  (\partial_{\bar z_1})^s(\partial_{\bar z_2})^{s'} \big[ \cM_{ss'}\tilde F ] \,,
\end{align}
where we introduced the shorthand notation 
\begin{align}\label{tilde-Phi}
\tilde F =  \int^\infty_{-\infty} dz_1^- dz_2^- F(u,v) =  { z_{12} \bar z_{12} z_3^- \over z_3^+}  \cG(\gamma)\,,
\end{align} 
and the function $\cG(\gamma)$ depends on the single variable
\begin{align}\label{gamma}
\gamma=-{(z_1\bar z_1-z_3^+z_3^-)(z_2\bar z_2-z_3^+z_3^-) \over  z_{12} \bar z_{12} z_3^+ z_3^-} \,.
\end{align}
To get the second relation in \re{tilde-Phi}, we used the expressions for
the conformal cross-ratios in the gauge \re{gauge1} and \re{gauge3}, 
\begin{align}
 u={z_{12}^2 z_{34}^2\over z_{13}^2 z_{24}^2}={z_{12}\bar z_{12} z_3^- \over (z_{13}^-z_3^+ + z_1\bar z_1)z_2^-} \,,\qqqquad 
v={z_{23}^2 z_{41}^2\over z_{13}^2 z_{24}^2} = {(z_{23}^-z_3^+ + z_2\bar z_2)z_1^- \over (z_{13}^-z_3^+ + z_1\bar z_1)z_2^-}\,,
\end{align}     
and rescaled the integration variables in \re{tilde-Phi} as 
\begin{align}
&z_1^-\to z_1^- (z_1\bar z_1-z_3^+z_3^-)/z_3^+\,,&&z_2^-\to z_2^- (z_2\bar z_2-z_3^+z_3^-)/z_3^+\,.  
\end{align}

For our purposes here we do not need the explicit expression for $\cG(\gamma)$ and refer the interested reader to \cite{Belitsky:2013xxa}.\footnote{We would like to point out that $F(u,v)$ describes the anomalous part of the \textit{Euclidean} correlation function \re{G4-loop}, while
the light-cone integrated correlation function \re{minus-int1} is an intrinsically \textit{Minkowskian} quantity. This means that we have to perform an analytic continuation of $F(u,v)$.  Knowing that $F(u,v)$ has complicated analytical properties, this can be a nontrivial task. 
Still, as was shown in Ref.~\cite{Belitsky:2013xxa,Belitsky:2013bja,Belitsky:2013ofa}, following~\cite{Mack:2009mi}, it can be easily done using the Mellin transform of the correlation 
function. }
It is more important to us that the same function $\cG(\gamma)$ appears in  \re{minus-int1} independently of the 
choice of currents in  \re{minus-int}
\begin{align} 
 G_{ss',{\rm top}}={z_{12} \bar z_{12}  z_3^+z_3^-( z_4^+)^2\over (n^+)^{s+1}(n'{}^+)^{s'+1}}  (\partial_{\bar z_1})^s(\partial_{\bar z_2})^{s'} \big[ \cM_{s s'}\cG(\gamma)] \,.
\end{align}
In the simplest case of scalar operators, $s=s'=0$ we obtain from \re{low-exp} and \re{G4-z}
\begin{align}\label{OO-2}
G_{00,\mathbf{20'}}=    { (y_1^2 y_4^2)^2 \over  n^+ n'{}^+}  
   { z_3^+   \over z_{12} \bar z_{12} z_3^- } \cG(\gamma)\,.
\end{align}
For the double insertion of energy-momentum tensors, for $s=s'=2$, we get from \re{TT-2}
 \begin{align}\label{TT-2}
G_{22,\mathbf{1}}=   {  (y_4^2)^2 \over (n^+)^{3}(n'{}^+)^{3}} 
   {(z_3^+)^3  \over (z_{12}   z_3^-)^3}  \lr{\partial_{\bar z_1} \partial_{\bar z_2}}^2 \big[    
 \bar z_{12}  \cG(\gamma)\big]  .
\end{align}    
Using \re{gamma}, we can cast this relation into the following form
\begin{align}\label{TT-3}
G_{22,\mathbf{1}}=   { (y_4^2)^2 \over (n^+)^{3}(n'{}^+)^{3}} 
   {(z_3^+)^3 \over (z_{12} \bar z_{12}  z_3^-)^3} \big[    
\gamma^2(1-\gamma)^2 \cG''(\gamma)\big]''  \,,
\end{align}
where the primes denote derivatives with respect to $\gamma$.
Repeating the same analysis for the remaining correlation functions \re{JO-TO}, \re{JJ-irr} and \re{TJ-top} we obtain
\begin{align}\notag\label{GGs}
& G_{10,\mathbf{175}} = {  y^{a'a}_2  y^2_2 (y_4^2)^2   \over (n^+)^{2}n'{}^+} \frac{ z_3^+  (z^+_3z^-_3 -z_2 \bar z_2)}{(z_3^- z_{12} \bar z_{12})^2}\ \left[(1-\g) \cG(\g)\right]'\,,
\\\notag
& G_{20,\mathbf{20'}} = { y^2_{2}(y_4^2)^2   \over (n^+)^{3} n'{}^+} \frac{ z_3^+  (z^+_3z^-_3 -z_2 \bar z_2)^2}{(z_3^- z_{12} \bar z_{12})^3}\  \left[(1-\g)^2 \cG(\g)\right]'' \,,
\\\notag
& G_{11,\mathbf{84}} = { (y_2)^{a'(a} (y_2)^{b'b)}(y_4^2)^2   \over (n^+)^{2}(n'{}^+)^{2}} \frac{(z_3^+ )^2 }{(z_3^- z_{12} \bar z_{12})^2} \ \left[\g(1-\g) \cG'(\g) \right]' \,,
\\
& G_{21,\mathbf{15}} =  {(y_2)^{a'a}(y_4^2)^2 \over (n^+)^{3}(n'{}^+)^{2}} \frac{ (z_3^+ )^2(z^+_3z_3^--z_2\bar{z}_2)}{(z_3^- z_{12} \bar z_{12})^3}\ \left[(1-\g)^2\g \cG'(\g)\right]''.
\end{align}
We recall that these relations were obtained in the gauge \re{gauge1} and \re{gauge3}. 
Examining the dependence of \re{GGs} on $s$ and $s'$ we observe the simple pattern
\begin{align}
G_{ss',\text{top}} \sim {d^s \over d\gamma^s} \left[(1-\gamma)^s \gamma^{s'}  {d^s \over d\gamma^{s'}}\cG(\gamma)\right]\,.
\end{align}
 
\subsection{Relations between charge flow correlations}

To compute the charge flow correlations \re{G-flow}, we have to restore the covariant form of \re{GGs}, revert from $z-$  to
$x-$coordinates using \re{5} and, finally, perform a Fourier transform of $G_{ss'}$  with respect to the the separation between the two scalar sources $x_{34}$.

The easiest way to do this is to apply the identities
\begin{align}\notag
&  
(x_{34} n) = {n^+\over 2z_3^+}  (z_3^+ z_3^-- z_1 \bar z_1)\,,  && x_{34}^2 = - {z_3^-\over z_3^+}\,, 
\\
&
(x_{34} n') = {n'{}^+\over 2z_3^+} (z_3^+ z_3^-- z_2 \bar z_2)\,,
&& 
 (n n') =\frac12 n^+ n'{}^+  z_{12} \bar z_{12} \,.
\end{align}
Here we changed the variables $x_{3,4}\to z_{3,4}$ according to \re{5} and replaced $\vec z_1=\vec n/n^+$ and $\vec z_2=\vec n'/n'{}^+$.  Using these relations we obtain the covariant form of \re{gamma}
\begin{align}
\gamma = {2(x_{34} n) (x_{34} n') \over x_{34}^2  (nn') }\,,
\end{align}
and express the factors entering $G_{ss'}$ in terms of $x_{34}^2$, $(x_{34} n)$, $(x_{34} n')$ and $(nn')$.
In a similar manner, we can undo the gauge \re{gauge1} and restore the dependence on $y_3$  in \re{GGs}
\begin{align}\notag\label{calY}
& y^{a'a}_2  y^2_2 (y_4^2)^2 \to y_{23}^2 y_{24}^2 (Y_{234})^{a'a}=\cY_{10,\mathbf{175}}\,,
&&
y_2^2 (y_4^2)^2 \to y_{23}^2 y_{34}^2 y_{42}^2=\cY_{20,\mathbf{20'}}\,,
\\[2mm]
& (y_2)^{a'(a} (y_2)^{b'b)}(y_4^2)^2 \to (Y_{234})^{a'(a}(Y_{234})^{b'b)}=\cY_{11,\mathbf{84}}\,,
&&
(y_2)^{a'a}(y_4^2)^2 \to y_{34}^2 (Y_{134})^{a'a}=\cY_{21,\mathbf{15}}\,,
\end{align}
where the $Y-$tensor was defined in \re{XY}.

Combining these relations, we obtain the covariant form of \re{OO-2}, \re{TT-3} and \re{GGs}
\begin{align}
 G_{ss',\text{top}} = \cY_{ss', \text{top}} { 2^{s'+1}(x_{34}n')^{s-s'} \over (nn')^{s+1}(x_{34}^2)^{s+1}}{d^s \over d\gamma^s}  \left[(1-\gamma)^s \gamma^{s'} 
 \cG^{(s')}(\gamma)\right],
\end{align} 
where $\cY_{ss', \text{top}}$ are the $y-$dependent structures defined in \re{calY}. It turns out that this relation admits another
representation
\begin{align}\label{Gss-new}
 G_{ss',\text{top}} = \cY_{ss', \text{top}} {1\over 2^{s-1} (nn')^{s+1} }(\Box_{x_3})^{s'}(n'\pa_{x_3})^{s-s'}  \left( \frac{\cG(\g)}{x_{34}^2} \right)\,,
 \end{align} 
which greatly facilitates  the task of computing the Fourier transform of $G_{ss'}$ in \p{G-flow}. We recall that the relation \p{Gss-new} is only valid for
the special contribution to the correlation function corresponding to the top Casimir channel in the $SU(4)$ decomposition \re{R-dec}.

Substituting \re{Gss-new} into the first relation in \re{G-flow} we finally obtain
\begin{align}\label{calF}
 \vev{\mathcal D_{s}(n) \mathcal D_{s'}(n')}_{q,\text{top}} =\cY_{ss', \text{top}} {(q^2)^{s'-1}(qn')^{s-s'} \over  2^{s-1}(nn')^{s+1} }  \cF (z)\,,
\end{align}
where $\cF (z)$ is the event shape function  introduced in \cite{Belitsky:2013bja}   
\begin{align}\label{FF}
&\cF (z) =q^2 \int d^4 x_{34} \e^{iq x_{34}}  \frac{\cG(\g)}{x_{34}^2-i0 x_{34}^0} \,,\qqqquad  z={q^2 (nn') \over 2(qn)(qn')} \,.
\end{align}     
Here the `$-i0 x_{34}^0$' prescription defines a particular analytic continuation of the Euclidean propagator $1/x_{34}^2$ to Minkowski space-time. Its choice is
dictated by the condition on the function $\cF(z)$ to take real values in the physical region $q^0>0$ and $q_\mu^2>0$, or equivalently
for $0< z< 1$.  

Comparing \re{calF} and  \re{F-def}, we obtain expression for the top Casimir component of the event shape function $\cF_{ss'} = \sum_R \cY_{ss';R} \cF_{ss',R}$ in terms of the function $\cF (z)$ defined in \p{FF}. The very fact that  $\cF (z)$ in \re{FF} does not depend on $s$ and $s'$  implies that the event shape functions for all charge flow correlations, restricted to the top Casimir channel,  coincide:
\begin{align}\label{univ}
 \cF_{00,\mathbf{20'}} (z) =\cF_{22,\mathbf{1}} (z) = \cF_{10,\mathbf{175}}(z)= \cF_{20,\mathbf{20'}}(z)= \cF_{11,\mathbf{84}}(z) = \cF_{21,\mathbf{15}}(z)\,.
\end{align}
This remarkable relation was discovered in \cite{Belitsky:2013bja} in the context of $\cN=4$ SYM, where it holds to all orders of perturbation theory as well as at strong coupling. As we explained in this section, this relation is not sensitive to the choice of the theory and 
follows from the $\cN=4$ superconformal symmetry of the four-point correlation function of the energy-momentum supermultiplet.         

\section{Conclusions}

In this paper we have developed a method for computing four-point correlation functions of the components of the $\mathcal{N}=4$ energy-momentum supermultiplet. 
Using the superconformal symmetry, we have written the four-point super-correlation function in the form of (\ref{G4-main}), with the generators of a maximal abelian  
fermionic subalgebra acting on Grassmann delta functions multiplying a scalar function of the conformal cross-ratios (dressed by a simple rational factor). While 
extracting a particular component of (\ref{G4-main}) is tedious (but straightforward) in general, we have discussed in great detail the subclass (\ref{hatG4}) of correlators, 
for which the computational effort can be significantly reduced. The correlators (\ref{hatG4}) are relevant for the computation of event shape functions in $\mathcal{N}=4$ 
superconformal theories, which were discussed in \cite{Hofman:2008ar} and recently worked out more systematically in \cite{Belitsky:2013xxa,Belitsky:2013bja,Belitsky:2013ofa}. 
Using the methods outlined in this paper, we have elucidated a number of interesting relations between different types of charge-flow correlations, which have first been 
noted in \cite{Belitsky:2013bja}.

A number of comments are in order concerning our work. First of all, our analysis solely relies on the $\mathcal{N}=4$ superconformal algebra and our results are 
therefore not specific to the dynamics of a particular theory (such as super-Yang Mills theory). The information about the latter enters only through the explicit form 
of the four-point function of the lowest half-BPS scalar operators of the energy-momentum supermultiplet, specifically through the function $\Phi(u,v)$, which was 
kept arbitrary throughout this work. This is true in particular also for the coupling constant of the theory and thus our results hold regardless of the order of perturbation 
theory. We therefore expect our work to be relevant for general treatments of $\mathcal{N}=4$ theories.

The class of correlation functions (\ref{hatG4}), which we discussed in the second part of the paper is a crucial ingredient in the computation of so-called charge-flow 
correlations in $\mathcal{N}=4$ theories. The intricate relations between different types of the latter point to some interesting structure, which appears to be a remnant 
of the original superconformal symmetry of the theory. Physically, some of the flow operators capture conserved observables and it would be interesting to study their 
relations with similar quantities in QCD. For a discussion at the two-loop level see~\cite{Belitsky:2013ofa}.

\section*{Acknowledgements}        

We are most grateful to A. Zhiboedov for collaboration on related topics. 
We would also like to thank J.~Drummond, B.~Eden, P.~Heslop, D.~Hofman, J. Maldacena, G.~Papathanasiou and R.~Stora for interesting discussions and conversations.       
 
 \appendix
 
 \section{Matrices and spinors}\label{App:A}
 \label{AppendixNotations}
 
 In this appendix we specify the spinor notations  that we use throughout the paper. 
 
 A four-dimensional vector $z^\mu=(z^0,\vec z)$ is represented by $2\times 2$ matrices  
 \begin{align}\label{35}
&z_{\a\da}=z_\mu(\bar \sigma^\mu)_{\a\da}= z^0 \s^0 +  \vec  z \cdot \vec \s = \left[\begin{array}{ll} z^+ & \bar z \\  z & z^-\end{array}\hspace*{-1mm}\right],
\nt
& z^{\da\a}=z_\mu( \sigma^\mu)^{\da\a}= z^0 \s^0 -  \vec  z \cdot \vec \s
= \left[\hspace*{-2mm}\begin{array}{cc} \phantom{-} z^- & -\bar z \\  -z &  \phantom{-} z^+\end{array}\hspace*{-1mm}\right]\,,
\end{align}
where $\sigma^\mu = (1, \vec{\sigma})$, $\bar\sigma^\mu = (1, - \vec\sigma)$ and $\vec \s =(\s^1,\s^2,\s^3)$ are the Pauli matrices.
Here we have introduced notation for the light-cone coordinates and the complex transverse components
\begin{align}\label{36}
z^\pm =z^0\pm z^3\,,\qquad
z= z^1+i z^2\,, \qquad \bar z = z^1-i z^2\, .
\end{align} 
It is easy to verify that
\begin{align}
z_{\a\da}z^{\da\b} =z_\mu z^\mu \delta_\a^\b
\,,\qquad 
z_\mu z^\mu 
=
\det \| z_{\a\da}\| =z^+ z^- - z \bar z\,.
\end{align} 
We adopt the following conventions for lowering and raising spinor indices
\begin{align}
&
\xi^\alpha = \epsilon^{\alpha\beta} \xi_\beta
\, , && \hspace*{-10mm}
\bar\xi^{\dot\alpha} = \bar\xi_{\dot\beta} \epsilon^{\dot\beta\dot\alpha}
\, , &&  \hspace*{-10mm} x^{\db\b} =  \epsilon^{\b\a}x_{\a\da}\epsilon^{\da\db} \,,
\end{align}
with Levi-Civita tensors normalized as 
\begin{align}
 \epsilon^{12} = \epsilon_{12} = -\epsilon_{\dot 1 \dot 2} =- \epsilon^{\dot 1 \dot 2} =1\,,
\end{align}
so that 
$\epsilon^{\alpha\beta}\epsilon_{\alpha\gamma}=\delta_\gamma^\beta$ and   $\epsilon_{\dot \alpha\dot \beta}\epsilon^{\dot \alpha\dot \gamma}=\delta^{\dot \gamma}_{\dot \beta}$. In particular,
\begin{align}
z^\a{}_\da  = \epsilon^{\a\b} z_{\b\da} =\left[\hspace*{-1mm}\begin{array}{cc}   z  & z^- \\  -z^+ &  -\bar z \phantom{-} \end{array}\hspace*{-2mm}\right],
\qqqquad
z_\a{}^\da =z_{\a\db}\epsilon^{\db\da}  
=
\left[\hspace*{-1mm}\begin{array}{cc}  \bar z  & -z^+ \\  \phantom{-}  z^- & - z \end{array}\hspace*{-1mm}\right]
\, .
\end{align}
For $SU(2)$ indices we adopt similar conventions for the components of the metric
\begin{align}
\epsilon^{12} = \epsilon_{12} = - \epsilon^{1'2'} = - \epsilon_{1'2'} = 1
\, .
\end{align}
Their raising/lowering is done according to the rules
\begin{align}
u^a = \epsilon^{ab} u_b
\, , && \hspace*{-10mm}
\bar{u}^{a'} = \bar{u}_{b'} \epsilon^{b'a'}
\, , &&   \hspace*{-10mm} y^{b' b} = \epsilon^{ba}y_{ a a'}\epsilon^{a' b'}\,. 
\end{align} 
The $SL(4)$ invariant intervals squared are introduced via the formula
\begin{align}
x^2 = - \ft12 x_{\alpha\dot\alpha} x_{\beta\dot\beta} \epsilon^{\alpha\beta}\epsilon^{\dot\alpha\dot\beta}\,,\qqqquad
y^2=-\ft12 y_{a'}{}^a y_{b'}{}^b \ep^{a'b'} \ep_{ab}
\end{align}
We use short-hand notation for the derivatives 
\begin{align}
(\pa_x)_{\a\da} 
= \frac{\partial}{\partial x^{\da\a}}
= \frac{1}{2} \sigma^\mu_{\a\da} \frac{\partial}{\partial x^\mu}
\, , \qqqquad 
(\pa_y)_{aa'} = \frac{\partial}{\partial y^{a'a}}
\end{align}
with simple action rule
\begin{align}
\label{x-tilde}
 (\pa_x)_{\a\da} x^{\db\b} =  \delta_\a^\b \delta_\da^\db\,,
 \qqqquad
(\pa_y)_{aa'} y^{b'b} = \delta_a^b \delta_{a'}^{b'}\,.
\end{align} 
Next, we define the projectors
\begin{align} 
\frac1{4}\s^- \bar\s^+ =\frac1{4} \bar\s^+ \s^-=\left(\begin{array}{cc} 0 & 0 \\ 0 & 1\end{array}\right)
\,, \qquad  
\frac1{4}\s^+ \bar\s^- =\frac1{4} \bar\s^- \s^+=\left(\begin{array}{cc} 1 & 0 \\ 0 & 0\end{array}\right)\,,
\end{align}
where $\s^+ = \bar\s^-= \s^0+\s^3$ and $\s^- = \bar\s^+= \s^0-\s^3$.
Representing the two-component spinors $\q_\a$ and $\bq^\da$ as columns, we decompose them with the help of projectors into
a sum of `$+$' and `$-$' components
\begin{align}\notag
\q_\a &=  \frac1{4}(\s^- \bar\s^+)_\a{}^\b \q_\b +\frac1{4}(\s^+ \bar\s^-)_\a{}^\b \q_\b =   
  \left(\begin{array}{l} 0  \\ \q^- \end{array}\right) + \left(\begin{array}{l} \q^+  \\ 0 \end{array}\right) ,
  \\
  \bq^{\da} &= \frac1{4}(\bar\s^- \s^+)^\da{}_\db \bq^\db   
  +\frac1{4}(\bar\s^+ \s^-)^\da{}_\db \bq^\db =  \left(\begin{array}{l} \bq^{-} \\ 0 \end{array}\right) + \left(\begin{array}{l} 0 \\ \bq^{+} \end{array}\right)  ,
\end{align}
so that $\theta^+ = \theta_{\alpha=1}$,  $\theta^-=\theta_{\alpha=2}$, and similarly for $\bar\theta^\pm$.
We also need spinors with lower/upper indices, which take the form of rows
\begin{align}\label{}
\q^\a = \epsilon^{\a\b} \q_\b = (\q^-,-\q^+)
\,,\qqqquad
\bq_\da = \ep_{\da\db} \bq^\db = (-\bq^{+},\bq^{-})\,.
\end{align}
   
\section{Harmonic superspace}\label{App:B}
\subsection{General description}
In this appendix we introduce our conventions for the harmonic variables which we use throughout the whole paper. They allow us to covariantly decompose any object in the fundamental representation of $SU(4)$, with quantum numbers of a particular subgroup. This can be done in two equivalent ways.

In the first case, the so-called {\it harmonic superspace}  approach to extended supersymmetry (see \cite{hh} for the $\cN=2$, \cite{Galperin:1984bu} for the $\cN=3$ and \cite{Hartwell:1994rp}  for  the $\cN=4$ versions), we introduce a harmonic matrix belonging to  $SU(4)$,
\begin{align}\label{su4}
u_B^A=(u^{+a}_B\,, \ u^{-a'}_B) \ \in \ SU(4)\,.
\end{align}
The lower index $B=1,2,3,4$ of this matrix transforms under the anti-fundamental representation of $SU(4)$. The upper index splits as $A=(+a,-a')$, according to the subgroup $SU(2)\times SU(2)' \times U(1) \subset SU(4)$, with indices $a,a'=1,2$ in the  fundamental representations of $SU(2)$ and $SU(2)'$, and a $U(1)$ charge $(\pm1)$, respectively. 

The $\cN=4$ harmonic variables defined in this way parametrize the four-dimensional complex compact coset
\begin{align}\label{cos2}
{\rm Gr}(4,2)\ = \  \frac{SU(4)}{SU(2)\times SU(2)' \times U(1)}\,,
\end{align}
which coincides with  the Grassmannian manifold ${\rm Gr}(4,2)$, i.e. the space of all two-dimensional linear subspaces of ${\mathbb C}^4$ \cite{Chern}.  

Using the harmonic variables (\ref{su4}), we can project the chiral odd coordinate of $\cN=4$ superspace $\q^A_\a$ onto two halves,
$\q^A_\a=(\q^{+a}_\a, \q^{-a'}_\a )$ with
\begin{align}\label{1.4}
   \q^{+a}_\a = \q^A_\a\, u^{+a}_A\,, \qqqquad \q^{-a'}_\a = \q^A_\a\, u^{-a'}_A \,.
\end{align}
The component $\q^{+a}_\a$ transforms as a doublet of $SU(2)$ and a singlet of $SU(2)'$ with $U(1)$ charge $(+1)$, and vice versa for the component
$\q^{-a'}_\a$. 
By maintaining the harmonic variables in the matrix form \p{su4}, we 
are able to do the decomposition \p{1.4} {\it without breaking $SU(4)$.}
 
In the second approach, the so-called  {\it analytic superspace} (for a review see \cite{Howe:1995md,Heslop:2003xu}), one complexifies the $R-$symmetry group, $SU(4) \to GL(4,{\mathbb C})$. Here the separation of $\q^A$ into two halves looks asymmetric,
$\q^A_\a =(\rho^a_\a\,, \vartheta^{a'}_\a)$ with
\begin{align}\label{1.10}
  \rho^a_\a = \q^a_\a + \q^{a'}_\a y_{a'}{}^a\,,\qqqquad
 \vartheta^{a'}_\a \equiv \q^{a'}_\a\,,
\end{align}
where $y_{a'}{}^b$ is a complex valued  $2\times2$ matrix. The decomposition \p{1.10}
corresponds to the alternative description of the Grassmannian ${\rm Gr}(4,2)$  \cite{Chern}:
\begin{align}\label{cos}
{\rm Gr}(4,2)\ = \ \frac{GL(4,{\mathbb C})}{{\cal P}}\,, 
\end{align}
where ${\cal P}$ is the parabolic subgroup of upper triangular matrices with $2\times2$ blocks. In Eq.~\p{1.10}, $\rho^a$ and $\vartheta^{a'}$ (with $a,a'=1,2$) transform under the subgroup $GL(2)\times GL(2)' \subset {\cal P}$ of the coset denominator. As usual with coset parametrizations, 4 of the (complex, i.e. not Hermitian) generators of  $GL(4)$ act transitively on the coordinates (as shifts of $y_{a'}{}^b$), another 8 act homogeneously (as $GL(2)\times GL(2)$ rotations of $y_{a'}{}^b$), and the rest are realized non-linearly.  The latter can be obtained by combining a shift of $y$ with the discrete operation of inversion, $y_{a'}{}^b \to y_{a'}{}^b/y^2$ 
(with $y^2=-\ft12 y_{a'}{}^a y_{b'}{}^b \ep^{a'b'} \ep_{ab}$), in close analogy with the action of the conformal group on the Minkowski space coordinates $x_{\a\da}$. Given two or more points, we can form covariant tensors, like $Y_{ijk}$ in \p{YI}.

The two equivalent descriptions of the manifold ${\rm Gr}(4,2)$ show two of its features \cite{Chern}. The harmonic description \p{cos2} makes its compactness manifest, the analytic description \p{cos} shows that it is holomorphic. In practice, to establish the relation between the two pictures, we replace the  unitary matrix $u$, Eq.~\p{su4},
and its hermitian conjugate $\bar u$ by a lower triangular  $GL(4)$ matrix and its inverse,
respectively,  
\begin{align}\label{3.17}
(u_B{}^{+a},\,  u_B{}^{-a'}) = \left(
\begin{array}{rc}
  \delta_{b}{}^a&   0  \\
  y_{b'}{}^a&    \delta_{b'}{}^{a'}
\end{array}
  \right)\,,\qquad
 (\bar u_{+a}{}^B,\, \ \bar u_{-a'}{}^B) =
  \left(
\begin{array}{rc}
  \delta_{a}{}^b&   0  \\
  -y_{a'}{}^b&    \delta_{a'}{}^{b'}
\end{array}
  \right)\,,
\end{align}
where $B=(+b,-b')$. In this notation \p{1.10} is  the equivalent of \p{1.4}, $ \q^{+a}_\a= \rho^a_\a$ and $\q^{-a'}_\a=\vartheta^{a'}_\a$.

\subsection{The special choice $y_1=y_2$}\label{slyy}

In Section \ref{s4} we  study detectors at points 1 and 2 and concentrate on  the special choice  $y_1=y_2$. It selects a particular $SU(4)$ channel in the tensor product, the one with the highest value of the quadratic Casimir (with the exception of the double-current insertion, see below). Here we give a simple explanation of this fact. 

For each choice of detectors (scalars, R-symmetry currents, energy-momentum tensors) the harmonic points $y_1$ and $y_2$ carry the corresponding representation of the R-symmetry group $SU(4)$ ($\mathbf{20'}$ for scalars, $\mathbf{15}$ for currents and singlet for energy-momentum tensors). Each of them is manifested as a particular representation of the little group $SU(2)\times SU(2)'\times U(1)$ of the harmonic coset, acting on the coordinates $y_1$ and $y_2$. When we identify the points $y_1=y_2$, we also identify these irreps. At the same time, the four-point function becomes a three-point function, built from the vectors $y_1=y_2$ , $y_3$ and $y_4$. This three-point function now carries the combined quantum numbers of $SU(2)\times SU(2)'\times U(1)$  at point $1\equiv 2$. Below we show that this corresponds (with the exception of two currents) to the irrep with top Casimir in the list of irreps for the particular tensor product. 

As the first example, consider two scalar detectors in the $\mathbf{20'}=[0,2,0]$.The first and third Dynkin labels measure the isospins of $SU(2)$ and $SU(2)'$, respectively, and the second label measures the $U(1)$ weight of the irrep. In the case at hand, there are no isospins, and the weights add up to form the irrep $\mathbf{105}=[0,4,0]$. This is the irrep of highest Casimir in the tensor product $\mathbf{20'}\times \mathbf{20'}$. 

The next example is a single current insertion at point 1. Now point 1 carries the irrep $\mathbf{15}=[1,0,1]$ while point 2 still carries $\mathbf{20'}=[0,2,0]$. In the limit $y_1=y_2$ the Dynkin labels add up to $\mathbf{175}=[1,2,1]$. This means that the resulting three-point function, made of the vectors $y_1=y_2$ , $y_3$ and $y_4$, carries isospins 1/2 of each $SU(2)$ group and weight 2. Again, this is the irrep of highest Casimir in the overlap of the  tensor products $\mathbf{15}\times \mathbf{20'}$ and $\mathbf{20'}\times \mathbf{20'}$.

Finally, the case with two currents is somewhat different. Now we have irreps  $\mathbf{15}=[1,0,1]$  at both points 1 and 2. After the identification, adding up the Dynkin labels we get  $\mathbf{84}=[2,0,2]$ (the top Casimir in the overlap of the  tensor products $\mathbf{15}\times \mathbf{15}$ and $\mathbf{20'}\times \mathbf{20'}$). However,  this is not the only possibility because the product of the two irreps of $SU(2)\times SU(2)'\times U(1)$ is reducible. Indeed, the two currents $J^{aa'}(y_1)$ and $J^{bb'}(y_2)$  transform under $SU(2)\times SU(2)'$ as well as under local $SU(4)$ inversions. This does not allow mixing the pair of indices $a,b$ (or $a',b'$). After the identification $y_1=y_2$ the  $SU(2)\times SU(2)'$ representation becomes reducible. To make it irreducible, we are now allowed to  (anti)symmetrize the  indices $a,b$ (and thus automatically $a',b'$): $J^{(a(a'} J^{b)b')}$ or $\ep_{ab}\ep_{a'b'} J^{aa'} J^{bb'}$. 
The symmetrization corresponds to adding up the first and third Dynkin labels (isospin $1/2\times 1/2 \to 1$) and we get the irrep $\mathbf{84}=[2,0,2]$. The antisymmetrization makes these two labels vanish (isospin $1/2\times 1/2 \to 0$) ; instead, the resulting singlet $y-$structure has $U(1)$ weight 2 (second Dynkin label), hence the irrep $\mathbf{20'}=[0,2,0]$.

\section{Invariance under $\bar Q$ and $S$ supersymmetry}\label{algebra}

In this appendix we give the detailed proof that $\cI_4$ from  \p{calI} is also invariant under $\bar Q$ and $S$ supersymmetry. To simplify the analysis it is convenient to introduce $SL(4)$ notation for the superalgebra generators. We combine $Q$ and $\bar S$ into an $SL(4)\times SL(4)$ matrix, $\cQ^M_A = (Q^\a_A, \bar S^\da_A)$, and similarly for $\bar Q$ and $S$, $\bar\cQ^A_M = (S^A_\a, \bar Q^A_\da)$. They satisfy the $\cN=4$ CSUSY algebra $SL(4|4)$
\begin{align}\label{al}
&\{\cQ^M_A,\bar\cQ^B_N\}=\delta^B_A L^M_N + R^B_A \delta^M_N\,, && \{\cQ^M_A,\cQ^N_B\}=\{\bar\cQ_M^A, \bar\cQ_N^B\}=0\,,
\nt [2mm]
& [L^M_N, \cQ^P_A]=-\delta^P_N \cQ^M_A + \frac1{4} \delta^M_N \cQ^P_C\,, &&  [L^M_N, \bar\cQ^A_P] = \delta^M_P \bar\cQ^A_N - \frac1{4} \delta^M_N \cQ^A_P\,, \nt  
& [R^B_A, \cQ^M_C]=\delta^B_C \cQ^M_A - \frac1{4} \delta^B_A \cQ^M_C\,, &&  [R^B_A, \bar\cQ^C_M] = -\delta^C_A \bar\cQ^B_M + \frac1{4} \delta^B_A \cQ^C_M\,.
\end{align}
Here $R^B_A$ and $L^N_M$ (with $R^A_A=L^M_M=0$) are the generators of two copies of $SL(4)$, the (complexified) R-symmetry and conformal groups, respectively. For the latter, restricting to various projections of the indices, we identify the familiar conformal group generators, for example $L^\a_\da \equiv P^\a_\da$, $L^\da_\a \equiv K^\da_\a$, $L^\a_\b= M^\a_\b + \delta^\a_\b D$, etc. 

In these terms the invariant \p{calI} can be rewritten more compactly,
\begin{align}\label{2.44}
&\cI_4 = (\cQ)^{16} \cA_4 \quad {\rm with} \quad  (\cQ)^{16} = \prod_{A,M=1}^4 \cQ^M_A  \,.
\end{align}
By construction, it is annihilated by $\cQ$ but it is not obvious why $\bar\cQ$ should also annihilate it.  This relies on the special properties of the function $\cA_4$ defined in \p{calA}. From the conjugated form of \p{generators} we see that all $\bar\cQ \sim a\pa_{\bq} + b\q\pa_x + c \q\pa_y$ therefore $\bar\cQ\cA_4=0$. What remains to show is that the commutator of $\bar\cQ$ with $(\cQ)^{16}$ in \p{2.44} is of the form 
\begin{align}\label{2.45}
[\bar\cQ^A_M,   (\cQ)^{16}] \sim \lr{(\cQ)^{15}}^B_N (\delta^A_B L^N_M + R^A_B \delta^N_M)\,.
\end{align}
Then, if $\cA_4$ is an R-symmetry and conformal invariant, $R\cA_4= L \cA_4=0$, we can conclude that $\bar\cQ\, \cI_4=0$. 

We can further simplify the analysis by replacing $(\cQ)^{16}$ with the following expression 
\begin{align}
 (\cQ\cdot \epsilon)^{16}  \sim (\cQ)^{16} (\ep)^{16}
\end{align}
where $\cQ\cdot \epsilon=\cQ_A^M \epsilon^A_M$ and $\epsilon^A_M$ is a $4\times4$ matrix of anticommuting variables. The algebra \p{al} yields
\begin{align}
[\bar \cQ_N^B, \cQ\cdot \epsilon] = \epsilon_M^B L_N^M +  \epsilon_N^A R_A^B\,, 
\end{align}          
hence
 \begin{align}
[\bar \cQ_N^B ,\,(\cQ\cdot \epsilon)^{16}] &=\sum_{k=0}^{15}(\cQ\cdot \epsilon)^{k}\, [\bar \cQ_N^B, \cQ\cdot \epsilon] \, (\cQ\cdot \epsilon)^{15-k}  =\sum_{k=0}^{15}(\cQ\cdot \epsilon)^{k}\, \big[ \epsilon_M^B L_N^M +  \epsilon_N^A R_A^B\big] \, (\cQ\cdot \epsilon)^{15-k} 
\nt
&= \sum_{k=0}^{14} (k+1) (\cQ\cdot \epsilon)^{k} \big[ \epsilon_M^B L_N^M +  \epsilon_N^A R_A^B, \cQ\cdot \epsilon\big]  (\cQ\cdot \epsilon)^{14-k} + 16  (\cQ\cdot \epsilon)^{15}(\epsilon_M^B L_N^M +  \epsilon_N^A R_A^B)\,.
\end{align}
The second term in the second line in this relation is of the expected form \p{2.45}. It remains to show that the first term vanishes. The commutation relations \p{al} yield\footnote{ This step uses the fact that the superalgebra is of the type $SL(m|n)$ with $m=n$. Otherwise on the right-hand side of \p{c9} there would be an extra term $\sim (1/m-1/n)\epsilon_N^B(\cQ\cdot \epsilon)$.}
\begin{align}\label{c9}
\big[ \epsilon_M^B L_N^M +  \epsilon_N^A R_A^B, \cQ\cdot \epsilon\big]= 2 \cQ_A^M \epsilon_M^B \epsilon_N^A \,.
\end{align}
Using the fact that $\cQ\cdot \epsilon$ commutes with $\cQ$ and with $\ep$, we find
\begin{align}\label{start}
\sum_{k=0}^{14} (k+1) (\cQ\cdot \epsilon)^{k} \big[ \epsilon_M^B L_N^M +  \epsilon_N^A R_A^B, \cQ\cdot \epsilon\big]  (\cQ\cdot \epsilon)^{14-k}= 240\, \cQ_A^M \epsilon_M^B \epsilon_N^A (\cQ\cdot \epsilon)^{14} \,.
\end{align}
Now we introduce vector notation, $\cQ^M_A = \cQ_m (\sigma_m)^M_A$ and  $\ep^A_M = \ep_m (\ts_m)^A_M$ (where $m=1,\ldots,16$), with the help of  $SO(16)$ chiral and antichiral sigma matrices obeying the Clifford algebra  $\sigma_m \ts_n+\sigma_n \ts_m= 2 \delta_{mn}\mathbb{I}$. In these terms the right-hand side of \p{start} becomes
\begin{align}\label{}
(\sigma_{m_1})^N_B \cQ_A^M \epsilon_M^B \epsilon_N^A (\cQ\cdot \epsilon)^{14}& \sim   (\cQ^{15})_{m_2} (\ep^{16}) \, \varepsilon_{m_1m_2 m_3 \ldots m_{16}} \varepsilon_{n_1n_2 m_3 \ldots m_{16}} \tr(\sigma_{m_1}\ts_{n_2} \sigma_{m_2} \ts_{n_1}) \nt [3mm]
&\sim (\cQ^{15})_{m_2} (\ep^{16}) (\delta_{m_1n_1}\delta_{m_2 n_2} -\delta_{m_1n_2}\delta_{m_2 n_1}) \tr(\sigma_{m_1}\ts_{n_2} \sigma_{m_2} \ts_{n_1})=0\,.
\end{align}

\section{Derivation of Eq.~\p{G4-main3}}\label{App:C}

In this appendix, we simplify the expression for the correlation function \p{G4-main1} in the gauge \p{gauge1} and \p{gauge3}.

To begin with, we observe that the product of generators $ \bar S^4\bar S'{}^4 Q^4Q'{}^4 $ in \re{G4-main1} is invariant under the 
transformations $\bar S\to \bar S+a\, Q$ and $\bar S'\to \bar S'+a' Q'$ with matrices $a$ and $a'$  independent of the coordinates at points 1 and 2. We can use this
freedom to define new generators
\begin{align}\label{cS}
\check S_{a\db} = \bar S_{a\db}  - (z_3)_{\a\db}  Q _{a}^\a   
\,,\qqqquad 
\check S_{b'\db} = \bar S_{b'\db}  -  (z_4)_{\a\db} Q _{b'}^\a   \,,
\end{align}
where $\bar S$ and $Q$ are given by the differential operators \re{generators}, with $x^\mu$ replaced by $z^\mu$. 
It is easy to check that, in the gauge \re{gauge1}, the generators $\check S$ and $\check S'$ defined in this way, do not involve derivatives with 
respect to $\theta_3$ and $\theta_4$. Therefore, evaluating \re{G4-main1} we can retain in $Q^4 Q'{}^4$
only terms containing the maximal number of derivatives with respect to $\theta_3$ and $\theta_4$, leading to
$Q^4 Q'{}^4 \lr{\theta_3^4 \theta_4^4} =  (y_4^2)^2$. Then, the expression in the second line of \re{G4-main1} can be 
simplified as
\begin{align} \label{sim1}  
\check S^4\check S'{}^4
\left[  (\q_1^+)^2   (\q_2^+)^2 {(\q_1^-)^2(\q_2^-)^2 \over (z_{13}^2 z_{24}^2 y_{1}^2)^2}{\Phi(u,v) \over uv}  \right]\bigg|_{\theta^+_{1,2}=\bar\theta_{1,2}^+=0}\,,
\end{align} 
where
$\check S$ and $\check S'$ are linear differential operators acting on the 
coordinates at  points $1$ and $2$
\begin{align}\notag\label{checkS}
& \check S_{a\dot\beta} = \sum_{i=1,2}   \left( z_{i3,\alpha\dot\beta}  {\partial \over \partial \q_{i,\alpha}^a}-\bq_{i,a'\dot\beta} {\partial \over \partial y_{i,a'}^a} \right),
\\
& \check  S_{b'\dot\beta} =\sum_{i=1,2}   \left( z_{i4,\alpha\dot\beta} \bq_{i,b'\dot\alpha} {\partial\over\partial z_{i,\alpha\dot\alpha}} +  z_{i4,\alpha\dot\beta} y_{i,b'}^a 
{\partial\over\partial \theta_{i,\alpha}^a}- \bq_{i,a'\dot\beta} y_{i,b'}^a  {\partial\over\partial y_{i,a'}^a} + \bq_{i,b'\dot\alpha} \bq_{i,a'\dot\beta}
{\partial\over\partial \bq_{i,a' \dot\alpha}} \right),
\end{align}
with $z_{i3} = z_i -z_3$ and similarly for $z_{i4}$.

Next, we define linear combinations of the generators \re{checkS}, which annihilate $\q_1^+$ and $\q_2^+$
\begin{align}\label{tildeS}
&\tilde  S_{a'\da} = \check  S_{a'\da}+ C^\db_\da{}^{a}_{a'} \check  S_{a\db}\,,\qquad\qquad\tilde  S_{a'\da} \theta_{1,2}^+ = 0\,.
\end{align}
To determine the  coefficients $C^\db_\da{}^{a}_{a'} $, we examine the action of $(\xi \cdot\tilde  S')=\xi_{\da a'} \tilde  S^{a'\da}$ on the Grassmann variables.
Taking into account \re{cS} and \re{CSUSY} we obtain
\begin{align}
\delta \theta_{i,\a}^a =(\xi \cdot\tilde  S')\,\theta_{i,\a}^a= \big[ (z_{i3})_{\a\db}C^\db_\da{}^{a}_{a'} + (z_{i4})_{\a\da}  (y_i)_{a'}^a\big] \xi^{\da a'}\,.
\end{align}
Recalling that $\theta_i^+=\theta_{i,\a=1}$, we impose the conditions $\delta \theta_{i}^+=0$ for $i=1,2$ to find
\begin{align}\label{xi}
 C^\db_\da{}^{a}_{a'} = { (z_{13})_{1\dot\gamma}  (z_{24})_{1\da}   (y_2)^a_{a'} - (z_{23})_{1\dot\gamma} (z_{14})_{1\da} (y_1)^a_{a'} \over  (z_{13} z_{23} )_{11}} \epsilon^{\dot\gamma\db}\,,
\end{align}
where we introduced the shorthand notation  $(z_{14})_{1\dot\alpha} \equiv (z_{14})_{\a=1,\dot\alpha}$ and similarly for the other matrix elements 
(see \re{35} in Appendix~\ref{App:A}). We can make use of the generators \re{tildeS} to write
\begin{align}\notag\label{sim2}
\check S^4\check S'{}^4  (\q_1^+)^2   (\q_2^+)^2  \big|_{\theta^+_{1,2}=0}
& =\check S^4\tilde S'{}^4 (\q_1^+)^2   (\q_2^+)^2 \big|_{\theta^+_{1,2}=0}
\\[2mm]
& =\left(\check S^4   (\q_1^+)^2   (\q_2^+)^2 \big|_{\theta^+_{1,2}=0}  \right) \tilde S'{}^4
=(z_{13}^+\bar z_{23} - z_{23}^+\bar z_{13})^2 \tilde S'{}^4\,.
\end{align}
In the second relation we replaced $\check S$ by its explicit
expression \re{checkS}, $\check S_{a\dot\beta} = \sum_{i=1,2}    (z_{i3})_{\alpha=1,\dot\beta} \partial_{\q_{i}^{a,+}}+\dots$ with
$  (z_{i3})_{\a=1,\db} =  (z_{i3}^+\,, \bar z_{i3})$.

Combining together \re{sim1} and \re{sim2} we finally obtain from \re{G4-main1}
\begin{align}\label{G4-main2}
& \vev{\cT_-(1) \cT_-(2)  \OO (3) \OO (4)} =   \lr{\frac{y^2_{1}}{\hat z^2_{13}}\frac{y^2_{4}}{\hat z^2_{24}}}^2\, (z_{13}^+\bar z_{23} - z_{23}^+\bar z_{13})^2\tilde S'{}^4
\left[    { (\q_1^-)^2 (\q_2^-)^2   \over (z_{13}^2 z_{24}^2 y_{1}^2 )^2}{\Phi(u,v) \over uv}  \right]\bigg|_{\bar\theta_{1,2}^+=0}\,,
\end{align}
where $\tilde S'$ is given by the differential operators \re{tildeS} and \re{checkS} acting on the coordinates at points $1$ and $2$.  
In particular, they involve derivatives with respect to the space-time coordinates $(z_i)_{\a\da}$ coming from the first term in the 
expression for $\check S_{b'\dot\beta}$. It is easy to see that for $\bar\theta_{1,2}^+=0$ this term involves derivatives with respect to 
$z_i^-$ and $\bar z_i$ only. The former derivatives produce vanishing contributions after the integration in 
\re{minus-int}, whereas the latter derivatives appear in the expression for the correlation function in the special form given in \re{short1}. 

This allows us to greatly simplify the expansion of \p{G4-main2} by dropping terms in the generators \re{tildeS} and 
\re{checkS} containing derivatives  $\partial/\partial y_i$ and $\partial/\partial \bq_i$ that do not yield such total derivatives. The result is 
\begin{align}\notag\label{tilde-S}
 \tilde  S_{a'\da}   = {}&  -(z_{14})_{1\dot\alpha} \bq_{1,a'}^- {\partial_{\bar z_1}}  -(z_{24})_{1\dot\alpha} \bq_{2,a'}^-  {\partial_{\bar z_2}}
 \\[2mm]  \notag
& + { (z_{23}z_{31}z_{14})_{1\da} (y_1)_{a'}^a + z_{13}^2 (z_{24})_{1\da}(y_2)_{a'}^a  \over (z_{13}z_{32})_{11}}{\partial_{\q_1^{a,-}}} 
 \\
& + { (z_{13}z_{32}z_{24})_{1\da} (y_2)_{a'}^a + z_{23}^2 (z_{14})_{1\da}(y_1)_{a'}^a  \over (z_{23}z_{31})_{11}}{\partial_{\q_2^{a,-}}} 
+ \dots\,.
\end{align}
This relation is valid for arbitrary space-time coordinates $z_i$. As in the previous section, we can significantly simplify
the calculation by imposing the gauge \p{gauge1} and \p{gauge3}.
 To evaluate \re{G4-main2} we introduce linear combinations of the generators
$\tilde  S_{a'\da}$ with $\da=(\dot 1, \dot 2)$
\begin{align} \label{rmS}
& {\rm S}_{1,a'} = z^+_3 \lr{ {\bar z_2 \over z^+_4} \tilde  S_{a'\dot 1} + \tilde  S_{a'\dot 2}}
\,,
\qqqquad {\rm S}_{2,a'} = z^+_3 \lr{ {\bar z_1 \over z^+_4} \tilde  S_{a'\dot 1} + \tilde  S_{a'\dot 2}}\,.
\end{align}
The explicit expression for these generators in the gauge \p{gauge1} and \p{gauge3} are given by \p{rmS1}.
The two generators in \re{rmS} are related to each other through the exchange of points $1$ and $2$.
 They anticommute with each other and satisfy  
\begin{align}
{\rm S}_{1}^2\, {\rm S}_{2}^2 = {(z_3^+)^4 (\bar z_{12})^2 \over (z_4^+)^2} \tilde S'{}^4\,,
\end{align}  
where ${\rm S}_{i}^2=\prod_{a'} {\rm S}_{i,a'}$. Combining this relation with \p{G4-main2} we arrive at \p{G4-main3}.
 

\end{document}